%% file: ngc891_paper2_r1.tex
\documentclass[useAMS,usegraphicx,usenatbib]{mn2e}

\usepackage{epstopdf}

\newcommand{{\apj}}{ApJ}
\newcommand{{\apjl}}{ApJLett}
\newcommand{{\apjs}}{ApJS}
\newcommand{{\aj}}{AJ}
\newcommand{{\mnras}}{MNRAS}
\newcommand{{\apss}}{Ap\&SS}
\newcommand{{\aap}}{A\&A}
\newcommand{{\aaps}}{A\&AS}
\newcommand{{\pasp}}{PASP}
\newcommand{{\aapr}}{ARA\&A}
\newcommand{{\araa}}{ARA\&A}
\newcommand{{\nat}}{Nature}

\title{The stellar population content of the thick disk and halo 
of the Milky Way analogue NGC 891}

\author[Rejkuba et al.]{M.~Rejkuba$^1$, M.~Mouhcine$^2$, R.~Ibata$^3$\\
$^{1}$ESO, Karl-Schwarzschild-Strasse 2, 
      D-85748 Garching, Germany \\
$^{2}$Astrophysics Research Institute, Liverpool John 
      Moores University, Twelve Quays House, Egerton 
      Wharf, Birkenhead, CH41 1LD, UK \\
$^{3}$Observatoire Astronomique de Strasbourg (UMR 7550),
      11, rue de l'Universit\'e, 67000 Strasbourg, France}

\date{Accepted ?. Received ?; in original form ?}

\pagerange{\pageref{firstpage}--\pageref{lastpage}}
\pubyear{2007}

\begin{document}

\maketitle

\label{firstpage}

\begin{abstract}

We present deep $VI$ images obtained with the Advanced Camera for Surveys 
on board the Hubble Space Telescope, covering three fields in the north-east 
side of the edge-on disk galaxy NGC~891. The observed fields span a wide range 
of galactocentric distances along the eastern minor axis, extending from the 
plane of the disk to 12~kpc, and out to $\sim25$~kpc along the major axis.
The photometry of individual stars reaches $\sim 2.5$ magnitudes below the 
tip of the red giant branch. We use the astrophotometric catalogue to probe 
the stellar content and metallicity distribution across the thick disk and 
spheroid of NGC~891. 

The colour-magnitude diagrams of thick disk and spheroid population are dominated 
by old red giant branch stars with a wide range of metallicities, from the sparsely 
populated metal-poor tail at $\mathrm{[Fe/H]}\sim -2.4$~dex, up to about half-solar 
metallicity. The peak of the metallicity distribution function of the thick disk 
is at $-0.9$~dex. The inner parts of the thick disk, within $\sim 14$~kpc 
along the major axis show no vertical colour/metallicity gradient. In the outer 
parts, a mild vertical gradient of 
$\Delta(V-I)_0/\Delta|Z| = 0.1 \pm 0.05$~kpc$^{-1}$, or less than
0.1~dex~kpc$^{-1}$ is detected, with bluer colours 
or more metal-poor stars at larger distances from the plane. This gradient is 
however accounted for by the mixing with the metal poor halo stars. No  
metallicity gradient along the major axis is present for thick disk stars, 
but strong variations of about $0.35$~dex
around the mean of $\mathrm{[Fe/H]}=-1.13$~dex are found. The properties 
of the asymmetric metallicity distribution functions of the thick disk stars show 
no significant changes in both the radial and the vertical directions. The stellar 
populations situated within the solar cylinder-like distances show strikingly 
different properties from those of the Galaxy populating similar distances. 
This suggests that the accretion histories of both galaxies have been different.

The spheroid population, composed of the inner spheroid and the halo, shows 
remarkably uniform stellar population properties. The median metallicity of the 
halo stellar population shows a shallow gradient from about $-1.15$~dex in the 
inner parts to $-1.27$~dex at 24 kpc distance from the centre, corresponding to 
$\sim 13$~r$_{eff}$. Similar to the thick disk stars, large variations around
the mean relation are present.
\end{abstract}


\section{Introduction}

The study of nearby spiral galaxies 
can provide an ``external view" of galaxies similar to the Milky Way (MW), and 
constrain the formation of spiral galaxy halos. The last decade has witnessed 
a spectacular increase in the discoveries of stellar streams and 
accretion events in the MW and Andromeda
\citep[e.g.][]{ibata+94,helmi+99,ibata+01,ferguson+02,ibata+03,yanny+03,martin+04},
showing that at least part of the stellar halos, and perhaps even disks,
of these galaxies are assembled from disrupted dwarf galaxies. 
However, the question whether the stellar halos in galaxies are built 
entirely through the multiple hierarchical merging
processes \citep[{\it a la}][]{searle+zinn78}, or rather have the majority of their 
mass assembled in an early dissipative collapse \citep[][]{ELS62} is still open
\citep[for the MW halo see the recent review by][]{helmi08}.
The thick disks in spiral galaxies, while observationally well established in most
galaxies \citep[e.g.][]{gilmore+reid83,dalcanton+bernstein02,seth+05b,mould05}, 
still need to be characterised better.
Through the observations of resolved stellar populations, it
is possible to determine the average metallicity,
metallicity distribution, radial profile and the amount of substructure.
These observables provide empirical constraints to the thick disk and 
halo formation models.

In contrast, the unambiguous detection of stellar halos in galaxies beyond the Local Group is
very challenging. The expected low metallicity and the low surface brightness make 
the surface photometry very challenging \citep{morrison+97,dejong08}. 
Extremely deep observations, and the technique of stacking of
huge number of similar edge on galaxies, produced detection of faint red halos
in distant galaxies \citep{zibetti+ferguson04,zibetti+04}.
However, the study of metallicity distributions of stars in 
individual galaxy halos
is only possible through resolved stellar photometry, which is limited by
current telescopes and instruments to about 10-12~Mpc. 
The emerging picture is that the inner regions of halos are relatively 
metal-rich with extended tails to lower metallicity 
\citep{durrell+01,durrell+04,mouhcine+05c,mouhcine06,mouhcine+07}. 
Selecting stars at very large distances, and ensuring that they show halo-like
kinematic signature, the metal-poor population, similar to the dominant component 
of the MW halo, was detected in M31 \citep{chapman+06,kalirai+06}. The halo of the 
MW has been shown to comprise two different stellar components that exhibit different 
chemical compositions, spatial distributions, and kinematics. 
This comparison between the two massive spirals of the Local Group shows the need 
for larger samples, and more detailed comparison with other spiral galaxies with 
similar structure and mass as massive galaxies in the Local Group, before general 
conclusions can be drawn about the formation and evolution of spiral galaxy halos.

Edge-on galaxies are the ideal targets, as this configuration permits a clear 
distinction between the halo, disk, and bulge, and allows for efficient search 
and characterisation of stellar substructures. With its high inclination 
\citep[$89.8\pm0.5$][]{kregel+vanderkruit05} NGC~891 is the ideal target for the 
study of vertical structure and stellar populations above the plane of the disk. 
At a distance of 9.7 Mpc \citep{mouhcine+07}, it is the closest edge-on spiral 
galaxy with morphological type, disk structure and mass similar to that of the 
Milky Way \citep{vanderkruit84, garcia-burillo+92}. However, it has been acknowledged 
in numerous studies, that it has much higher vertical extent of neutral and ionised 
gas structure \citep{garcia-burillo+92,scoville+93,kamphuis+07}, with a huge neutral 
HI halo surrounding it \citep{oosterloo+07,sancisi+08}. 

\citet{seth+05a} used ACS on board HST to resolve the stars in a field centred
on the thin disk of NGC~891. However, the crowded inner regions observed, 
and the shallow images did not allow them to study the stellar population 
properties of the observed field in more detail. The deeper ACS images (one of 
the three fields presented in this work) allowed \citet{tikhonov+05} to resolve 
red giant branch (RGB) stars, determine the distance from the luminosity of the 
RGB tip, and discuss the RGB and asymptotic giant branch (AGB) stellar distribution
along the minor axis. 

In a series of papers we study in detail the structure, metallicity distribution, 
and stellar populations of NGC~891 using archival observations of three 
HST/ACS fields. In the first paper \citep[][Paper~I]{mouhcine+07} we investigated 
the metallicity distribution of the stellar halo of NGC~891. The results of that 
study revealed a surprisingly high average metallicity, of 
$\mathrm{[Fe/H]}\approx -0.9$~dex at 9.5~kpc above the galaxy disk. 
This is $\sim 0.5$ dex more metal-rich than the Milky Way halo 
\citep[e.g.][]{carollo+07,ivezic+08}. \citet[][Paper~III]{ibata+09} investigated 
the structure of the halo and the disk of this galaxy. Using starcounts which cover
much larger area than in the work of \citet{tikhonov+05}, we detected the thick disk 
with vertical scale height of $h_Z=1.44 \pm 0.03$~kpc, and radial scale length of 
$h_R=4.8\pm 0.1$~kpc. Moreover, in the stellar spheroid, which is well fitted with 
a de Vaucouleurs-like profile out to the edge of our survey at $r\sim 25$~kpc, 
significant small-scale variations in the median colour and density are detected  
over the halo area. \citet[][Paper~IV]{harris+09} investigated the globular cluster population 
in the observed fields.

In this paper we present in detail the complete dataset and data reduction, the 
photometric catalogue, and completeness simulations. 
The colour-magnitude diagrams and metallicity distribution of stars are used to 
investigate the stellar content of the thick disk and halo of NGC~891.

\section{Data and photometry}
\label{sect:data}


\begin{figure}
\resizebox{\hsize}{!}{
\includegraphics[angle=0]{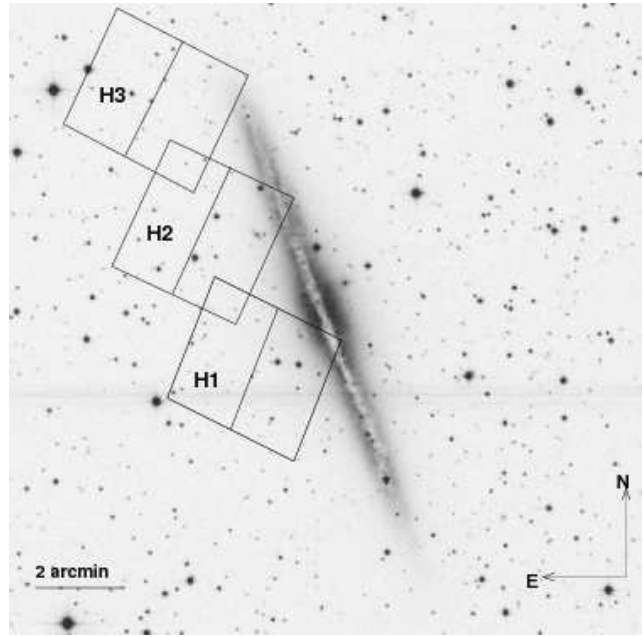}}
\caption{HST/ACS fields H1, H2, and H3 overlayed on 
Digitized Sky Survey image of NGC~891. The scale bar
of 2\arcmin\, corresponds to a physical scale of 5.7~kpc.}
\label{N891_fields}
\end{figure}

\begin{figure}
\resizebox{\hsize}{!}{
\includegraphics[angle=0]{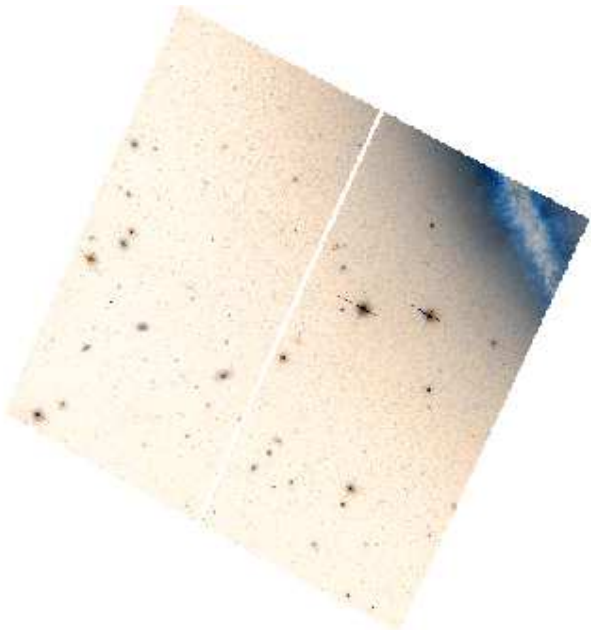}}
\caption{H1 field: this image is the deep combination of
9 exposures in F606W (blue) and 9 exposures in F814W (red). 
The colours show the presence of young blue stars along the edge of the disk, 
and a redder population in the halo. The few very bright stars are 
foreground Milky Way stars, and several
background galaxies are visible through the halo of NGC 891.}
\label{N891_f1deep}
\end{figure}


\input{ngc891_paper2_tab01}

Observations of three fields (H1, H2, and H3) along the 
eastern edge of NGC~891 disk (Fig.~\ref{N891_fields}) have
been taken in February 2003 with the wide field camera (WFC) of ACS 
as part of programme GO-9414. The field of view of the ACS WFC 
is 202 square arcsec and the pixel size is 0.05 arcsec. At the distance
of NGC~891 (9.7 Mpc) the field of view covers 90 kpc$^2$, and the pixels 
are 2.4 pc in size.

Each field was observed with 9 $F606W$ exposures and 9 $F814W$ exposures.
These HST filters, similar to ground based $V$ and $I_c$ bands, 
are often used in studies of resolved stellar populations 
in the halos of galaxies, where most of the sources are expected to be red giants, 
because of the match in the sensitivity of the filter+detector response and 
the emission of the sources. 
The observing strategy was such that for each field three sets of three 
exposures per filter were taken. These three triplets were 
dithered with sub-pixel shifts between them. 
Each triplet of images was taken to ensure cosmic ray rejection 
through the CR-split observing strategy with exposure times tuned to 
fit the three exposures within one orbit.
The summary of the observations, with the coordinates of the centre of
the pointings, exposure times, and observing dates are given in
Table~\ref{table:logs}. The total integration time per filter per field is
7710 sec.

We used the flatfielded (\_flt), 
cosmic-ray rejected (\_crj), and drizzled (\_drz) images produced by the
``on-the-fly" reduction pipeline of the ESO/ST-ECF archive.  
The drizzled images obtained from the archive had combinations of only 3 CR-split 
exposures for each dither position. Therefore to make the deepest possible 
stack for each field, using the full 7710 sec exposures we run the
{\it multidrizzle} \citep{koekemoer02} task within the {\it stsdas} 
package in pyraf  to combine the 9 images
taken for each filter, separately for the three fields and two filters.
In Fig.~\ref{N891_f1deep} we show the combined deep image of the H1 field.

\subsection{Photometry}
\label{sec:dolphot}


\begin{figure}
\resizebox{\hsize}{!}{
\includegraphics[angle=0]{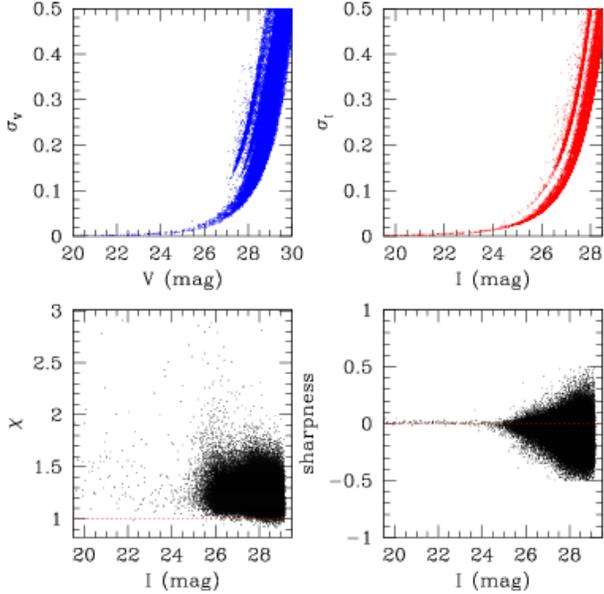}}
\caption{Photometric quality parameters for the H3 field.}
\label{fig:seleh3}
\end{figure}

\begin{figure}
\resizebox{\hsize}{!}{
\includegraphics[angle=0]{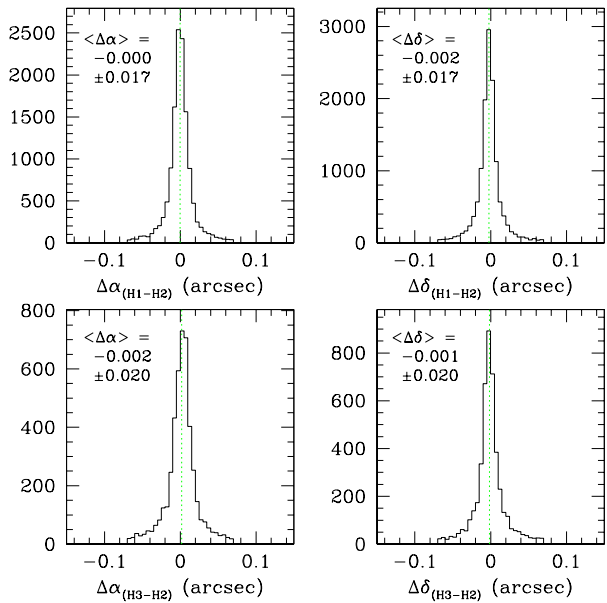}}
\caption{Offsets between the RA and DEC coordinates for objects detected in
common in the overlap regions H1-H2 (top) and H2-H3
(bottom).}
\label{fig:overlapdist}
\end{figure}

\begin{figure}
\resizebox{\hsize}{!}{
\includegraphics[angle=0]{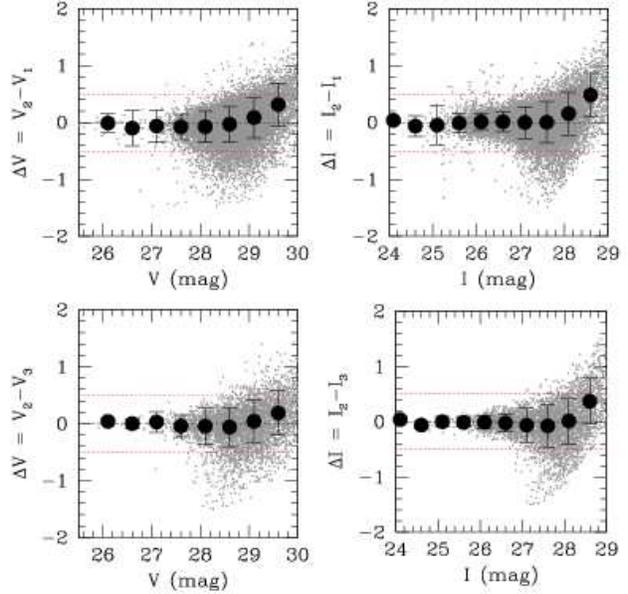}}
\caption{$V$ and $I$ band magnitude differences in the overlap regions. Top
panels: comparison between H1 and H2 field magnitudes for matched stars. 
Bottom panels: comparison between H2 and H3
field  photometry for matched stars.}
\label{fig:overlap}
\end{figure}

Due to large geometrical distortions of the ACS field of view the pixel size 
varies across the field resulting in incorrect photometry for point sources 
in flat-fielded (\_flt) images. The drizzle, and multidrizzle packages have been
developed to correct these distortions. However, the photometry done on the 
\_drz images is non-optimal because the stars in different parts of the field 
of view are resampled in a different way by the drizzling process. Moreover, the
signal in the adjacent pixels is correlated.
Therefore following the recommendations of \citet{anderson06} and the DOLPHOT/ACS 
User's Guide \citep{dolphin05} we decided to run the PSF fitting on the images that were not
corrected for distortion. To obtain correct flux measurements for point sources 
these images need to be multiplied by a pixel area map (PAM) prior to PSF
fitting. PAM files are available for 
all the filters from the Space Telescope Science Institute (STScI) web site. 

PSF fitting photometry of all the stellar objects in the images 
was run with the DOLPHOT photometric package, which is an
extension of HSTphot \citep{HSTphot}, and contains the specialised ACS
module tailored for accurate PSF photometry of ACS images. It contains
image preparation and processing routines that should be run prior to the 
 {\it dolphot} PSF
fitting programme in the following order. 
The first step is running the
{\it acsmask} programme in order to mask out all the 
pixels flagged as bad from the data quality extension of the images, and to 
multiply the images with PAM files and so obtain geometrically corrected 
images in units of electrons. The next step of data preparation is sky 
measurement using the {\it calcsky} routine. 
Finally, the alignment between the reference image and the input images is 
calculated using {\it acsfitdistort} programme. The psf fitting programme 
{\it dolphot} is run with the {\tt UseACS} flag set to 1 and the parameter file 
with the recommended values for the PSF radius, aperture size, sky region,
and threshold for detections. 
The psf fitting is run simultaneously on the 
set of input images, where the coordinates for the fitting objects are
derived from the reference image. All the parameters for the fitting, as 
well as the list of input images are passed through a configuration file. 

Given the small overlap between the three fields, we run the photometry 
separately for each of the three fields. The stars detected in common 
in the overlap regions were used to verify the accuracy of the photometry 
(see below).

We have made several runs of DOLPHOT photometry: 
the first run used as the reference image one of the \_crj images 
provided by the pipeline,
which is the combination of only 3 out of 9 exposures for a given field, and
as input data all the available \_crj images. The second run was done 
using as the reference again single \_crj image, but with \_flt 
images in input, where each \_flt image was an average of three CR-split exposures
taken at the same offset position and with the same filter.  
Finally, in the last run we used for the reference the 
deepest exposure, which we created ourselves using the {\it multidrizzle} 
task within pyraf. It consisted of the combination of  all the 9 exposures 
taken with the $F814W$ filter into a single deep image per field.

\begin{table*}
\caption{astrophotometric catalogue}
\label{table:cat}
\begin{tabular}{cccccccccccccccc}
\hline
(1)  &   (2)  &   (3)  &    (4)    &     (5)     &   (6)  &     (7)      & (8)   & (9)&(10)   & (11)  &  (12)  &  (13) &  (14)  &(15)  &  (16) \\
ID   &   RA   &   DEC  &   V(mag)  & $\sigma_V$  & I(mag) &  $\sigma_I$  & $\chi$&S/N &sharp  & round &  crowd &  typ  &  f     &vflag &  iflag\\
\hline
1   &     35.646270 & 42.308975  & 21.212  & 0.002 &   19.155  & 0.003  &  1.89  &	1385.3 &  0.017  &  0.027  &  0.061  &  1 &	  1    & 4 &	   2\\
2   &     35.671210  & 42.308643  & 21.086  & 0.002 &   19.522  & 0.002  &  1.93  &	1371.3 &  0.045  &  0.007  &  0.025  &  1 &	  1    & 4 &	   6\\
3   &     35.654569  & 42.308503  & 21.011  & 0.002 &   19.595  & 0.002  &  1.28  &	1362.0 &  0.009  &  0.010  &  0.019  &  1 &	  1    & 6 &	   4\\
4   &     35.664751  & 42.336123  & 21.482  & 0.002 &   19.298  & 0.003  &  1.41  &	1404.6 &  0.015  &  0.013  &  0.002  &  1 &	  1    & 4 &	   4\\
5   &     35.675835  & 42.303078  & 20.990  & 0.002 &   19.766  & 0.002  &  1.83  &	1358.2 &  0.050  &  0.041  &  0.001  &  1 &	  1    & 4 &	   4\\
\hline
\end{tabular}
\end{table*}

The difference between the first two photometry runs allows one to evaluate how
successful DOLPHOT is in rejecting the numerous cosmic rays when using \_flt
images in input which are not cosmic-ray cleaned. The last run was then used to
obtain the deepest photometry.
The photometry from all these runs provides the same results for the
bright part of the stellar populations present in the images. 
We note that the photometry
used in our first analysis of the halo metallicity distribution  in 
NGC~891 \citep{mouhcine+07} was based on \_crj reference images. 
Here we present the results obtained using the deepest multidrizzled images as
reference image.

The output of DOLPHOT contains the instrumental magnitudes as well as 
transformed, calibrated magnitudes for all the fitted objects. The 
transformations have been made using calibrations from \citet{sirianni+05} 
and include aperture corrections, as well as CTE loss corrections following
\citet{ACSCTE}. For each object the global  solution is listed first, 
and then the
photometry results are given for each of the input images. Here we use the 
global, combined photometry. Together with the magnitudes and the associated 
errors DOLPHOT provides also a range of quality flags: $\chi$, 
signal-to-noise, sharpness, roundness, crowding, ellipticity, and 
object type. These flags were used to select the bona fide stellar objects. 
Our selection criteria were the following: 
(1) object type 1 or 2, corresponding to a stellar (point source) object; 
(2) $\chi < 3$; 
(3) sharpness between $-0.5$ and $0.5$; 
(4) crowding parameter smaller than 0.35;
and (5) detection in both $V$ and $I$ bands with global photometric errors
smaller than 0.5 mag.

Applying these selection criteria, the photometric catalogues contain 
149076 stars in H1,  138133 stars in H2, and 106647 stars 
in the H3 field. 

In Fig.~\ref{fig:seleh3} the photometric quality parameters for the H3
sources are shown: magnitude error ($\sigma$), sharpness, 
and $\chi$ of the PSF fit, as a function of magnitude. Magnitude error values ($\sigma$)
in Fig.~\ref{fig:seleh3} are DOLPHOT values multiplied by $1.6$ to account for the fact
that the photometry was run combined averages of 3 \_flt images, while DOLPHOT computed 
the photometric errors based on the noise characteristics of the raw ACS images. 
Although in principle this correction factor should have been $\sqrt{3}$, 
we adopted the value of 1.6 based on the comparison of the photometry in the overlapping 
regions between fields (see Sect.~\ref{completeness} for details). There are several parallel
sequences in the magnitude-error plots. The larger errors at a given magnitude are
assigned to stars that are not detected in all the \_flt images, but only in a subset 
of them. The sharpness parameter for a perfect star has value of 0, it is negative for 
sharper, more spiky objects, possibly contaminated by some remaining cosmic 
rays, or bad pixels, and has positive values for more extended objects. 
The $\chi$ parameter is a measure of how well the model PSF matches the 
light distribution of the star. 
The crowding parameter measures the change in brightness for the star if the 
neighbours are not subtracted, and is expressed in magnitudes. It is an
indicator of how much blending there is due to overlapping PSF wings of
neighbouring stars.

The final photometric catalogue contains all the stars detected in the three
fields. The H2 field overlaps with both H1 and H3 in its corners.
To make the final catalogue we derived RA and DEC coordinates for all the
stars using pyraf task {\it xytosky}.  The relative accuracy of 
the astrometry was checked by overplotting the photometric catalogue over 
the \_drz images. Several bright stars were identified in 
common between the two independent catalogues in overlap regions, and these stars
were used to compute the initial shifts between H1 and H3, with respect to H2 field. 
Based on coordinates and $VI$ magnitude matches of all the stars these shifts 
were refined iteratively, and the stars detected independently in two fields 
in the overlap regions were found. 
There are 13054 and 4534 detections in H1-H2, and H2-H3 overlap 
regions, respectively. Fig.~\ref{fig:overlapdist} shows
the distribution of differences in RA and DEC between the 
stars found in the overlap regions. The differences in magnitudes for these
stars are shown in Fig.~\ref{fig:overlap}. There is
no systematic offset between H2 and H1/H3 detections for stars above the 50\%
completeness limits. 
In the final catalogue we use the average magnitude measured independently on two 
different sets of images for these matched stars.

Our final photometric catalogue contains 377320 stars detected in both $F606W$
and $F814W$ images in the three fields. The first 5 lines of the catalogue are
given in Table~\ref{table:cat}. The full catalogue is available in the electronic
version. The columns of the catalogue are: (1) id number; (2) and (3) RA 
and DEC in degrees; (4)--(7) calibrated V and I-band magnitudes and the 
associated errors as computed by DOLPHOT, but multiplied by a factor of $1.6$ to 
correct for the underestimated DOLPHOT errors (see below for details); 
(8) $\chi$ value of the PSF fit;
(9) signal-to-noise; (10) sharpness; (11) roundness; (12) crowding; 
(13) object type; (14) field (H1, H2, H3 or overlap 12 or 23); 
(15) and (16) V-band and I-band flags. These last flags give an indication of whether 
a given star had some bad or saturated pixels within the PSF fitting radius
(see the DOLPHOT manual by \citet{dolphin05} for details).

We have defined the following coordinate system with respect to the
major and minor axis of the galaxy: the centre of NGC~891 was taken to lie at
${\rm RA}_{\circ} = 02^{\rm h}22^{\rm m}33\fs4$, 
${\rm Dec}_{\circ} = 42\degr 20\arcmin 57\arcsec$
(J2000.0), and the position angle of the major axis 
was $\rm PA=22\degr$. In this coordinate system positive $X$ 
is located Northeast of the centre of NGC~891, and negative $Z$ is  
to the East. The ACS images cover $\sim 12$ kpc perpendicular to 
the disk, and $\sim 25$ kpc along the major axis.

\subsection{Error analysis and completeness}
\label{completeness}

\begin{table}
\caption{Photometric errors.}
\label{tab:errfits}
\begin{tabular}{r|rrr|rrr}
\hline
               & \multicolumn{3}{c}{F606W}& \multicolumn{3}{c}{F814W}\\
$\frac{N_{RGB}}{\mathrm{kpc}^2}$ &   $c_1$   &   $c_2$ &   $c_3$& $c_1$   &   $c_2$ &   $c_3$  \\
\hline
$450-1200$ &2.54E-7& 0.47&  0.18&   5.25E-7 &  0.52& -1.18	 \\
$140-450$  &3.37E-8& 0.51&  0.87&   9.25E-7 &  0.63& -5.26  \\
$44-140$   &1.17E-9& 0.70& -1.09&   2.17E-10 & 0.74&  0.16  \\
$<44$       &2.13E-10& 0.76& -0.94&  5.18E-10 & 0.74& -0.73  \\
\hline
\end{tabular}
\end{table}

\begin{table}
\caption{Completeness parameters.}
\label{tab:complfits}
\begin{tabular}{r|rr|rr}
\hline
               & \multicolumn{2}{c}{F606W}& \multicolumn{2}{c}{F814W}\\
$\frac{N_{RGB}}{\mathrm{kpc}^2}$ &   $m_0$   &   $\alpha$ & $m_0$ & $\alpha$  \\
\hline
$>1200$         & 27.56 & 1.55  &  26.15  & 1.2 \\
$450-1200$ & 28.40 & 1.2	&  27.33  & 1.0 \\
$140-450$  & 28.82 & 1.2	&  27.82  & 1.0 \\
$44-140$   & 29.02 & 1.2	&  28.13  & 1.1 \\
$<44$       & 29.03 & 1.2	&  28.15  & 1.1  \\
\hline
               & \multicolumn{2}{c}{F606W}& \multicolumn{2}{c}{F814W}\\
Z (kpc)   &   $m_0$   &   $\alpha$ & $m_0$ & $\alpha$  \\
\hline
$-2$ to $+1$   & 27.2  & -       &  25.8    & -    \\
$-4$ to $-2$   & 28.4  &  1.0    &  27.27 & 0.7 \\
$-6$ to $-4$   & 28.9  &  1.2    &  27.9   & 1.0 \\
$-9$ to $-6$   & 29.02 &  1.2    &  28.1   & 1.0 \\
$-12$ to $-9$  & 29.1  &  1.1    &  28.22 & 0.9 \\
\hline
\end{tabular}
\end{table}

Completeness simulations were run for all three observed fields. Artificial
star lists with at least 100,000 stars per field were created using {\it acsfakelist} programme
within DOLPHOT, and then DOLPHOT was run using the {\tt FakeStars} parameter in the
configuration file equal to the artificial star list. As DOLPHOT adds one fake
star to the image at the time and then re-measures its photometry, there is no
danger of creating ``overcrowded" image by adding too many stars with
overlapping PSF wings. In the output file of the completeness photometry run
all the stars, including those that were added to the images, but not detected,
are listed. Therefore it is easy to compute the completeness ratio: the number
of detected divided by the number of added fake stars. The criteria to detect
stars were set equal to those for selection of ``good" stars in the original
photometry runs, by selecting only stars that satisfy all the $\chi$, sharpness,
crowding, and magnitude error cuts. 


\begin{figure}
\resizebox{\hsize}{!}{
\includegraphics[angle=270]{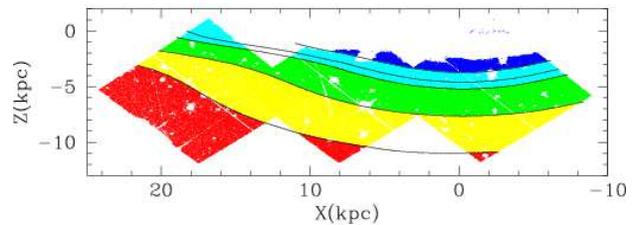}}
\caption{Distribution of the observed stars in the three ACS fields are plotted in the
coordinate system of NGC~891. Different colours denote regions of different density of RGB
stars according to interpolated observed  density profile \citep{ibata+09}. The number of
RGB giants per square kiloparsec in these regions are as follows: 
region 1 (blue) $N(RGB)/\mathrm{kpc^2} > 1200$,
region 2 (cyan)  $1200> N(RGB)/\mathrm{kpc^2} > 450$,
region 3 (green) $450> N(RGB)/\mathrm{kpc^2} > 140$,
region 4 (yellow) $140> N(RGB)/\mathrm{kpc^2} > 44$,
region 5 (red) $44> N(RGB)/\mathrm{kpc^2}$. The black
lines indicate the limits of the regions and are used to select the same regions (in X-Z
space) from the completeness simulations. An additional black line overplotted on region 2 (cyan) indicates the limit of the stellar density of $\sim 650$ $N(RGB)/\mathrm{kpc^2}$.}
\label{fig:density_regions}
\end{figure}

\begin{figure}
\resizebox{\hsize}{!}{
\includegraphics[angle=0]{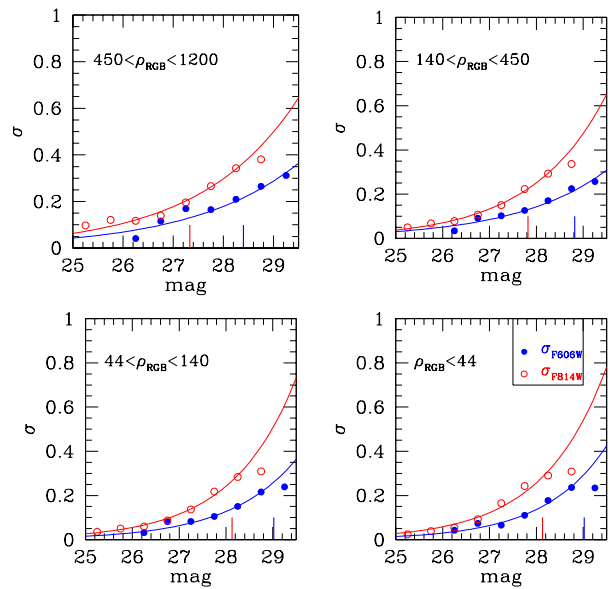}}
\caption{Average error as a function of magnitude and stellar density in the field
based on completeness simulations. The errors are shown separately for
each region (see Fig.~\ref{fig:density_regions}) selected according to the density of RGB
giants observed across the field. Solid lines are the fits of  the analytic function
given in Equ.~\ref{eq:errorfunc}. The coefficients of the fits are given 
in Table~\ref{tab:errfits}.}
\label{fig:magerr_dens}
\end{figure}

\begin{figure}
\resizebox{\hsize}{!}{
\includegraphics[angle=270]{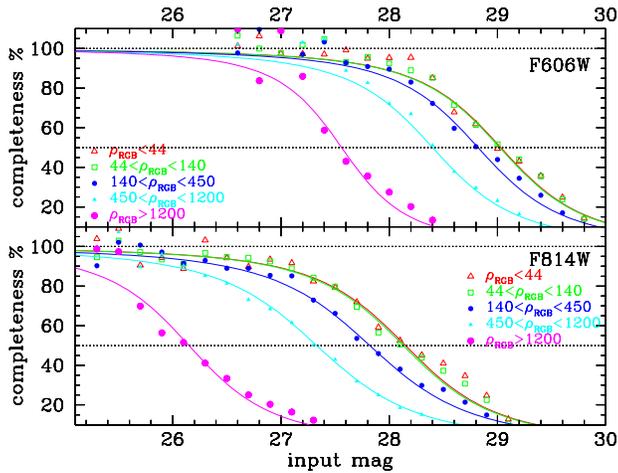}}
\caption{Completeness for $F606W$ (upper panel) and $F814W$ (lower panel) 
bands as a function of magnitude. Different symbols
are used to plot completeness functions for different regions according to the density of 
RGB stars (see Fig.~\ref{fig:density_regions}), and
the selected ranges are given in the diagrams. The lines are fitted analytical functions
of the form Equ.~\ref{eq:complfitequ} \citep{fleming+95}. The coefficients of the fits are given 
in Table~\ref{tab:complfits}.}
\label{fig:compldens}
\end{figure}

Given the large gradient in stellar density across the observed fields, we 
investigated the dependence of completeness and magnitude measurement errors as a
function of distance from the plane and stellar density. In Figure~\ref{fig:density_regions}
we show the distribution of all the observed stars colour-coded according to the number
density of RGB stars ($\rho = N(RGB)/\mathrm{kpc^2} $). We use here the interpolated density profiles 
derived in Paper~III. Solid black lines delimit 5 regions that have the 
following range of RGB star densities: 
region 1 (blue) $\rho > 1200$,
region 2 (cyan)  $1200> \rho> 450$,
region 3 (green) $450> \rho> 140$,
region 4 (yellow) $140> \rho> 44$,
region 5 (red) $44> \rho$. 
These same regions, selected based on X, Z galactocentric
coordinates were used to derive completeness and magnitude error dependence on magnitude in
Figures~\ref{fig:magerr_dens} and \ref{fig:compldens}.

By comparing directly the errors from DOLPHOT and from
magnitude differences in the overlap regions it is possible to verify the photometric
errors computed by DOLPHOT.
The overlap region between the H2 and H3 fields coincides with region 4 (RGB number density
between 44 and 140 stars), while the overlap between H2 and H1 has most of its stars
within region 3, and some in region 2. 
The ratio between DOLPHOT errors and the average scatter of the magnitude differences as a
function of magnitude is computed using the following expression:
\begin{equation}
\mathrm{ratio} = \frac{\mathrm{mag}_{H2} - \mathrm{mag}_{H3}}{\sqrt{\sigma_{H2}^2 + \sigma_{H3}^2}}
\label{eq:ratio_overlap_dolphot}
\end{equation}
where $\sigma_{H2}$ and $\sigma_{H3}$ are the photometric errors for stars obtained from DOLPHOT in the corresponding field.
This ratio indicates that DOLPHOT underestimates the errors by as much as a factor of 1.6. This
factor is consistent with the expected factor of $\sqrt{3} = 1.7$, which is due to the fact that the input images are averages of three \_flt frames, while 
DOLPHOT computed the magnitude errors using the noise characteristics of raw images.
All the uncertainties in Table~\ref{table:cat} incorporate this factor of 1.6, and are
adopted in Paper III and in all subsequent analysis in this contribution.

The dependence of the photometric error measurements on
magnitude for each region is given in Figure~\ref{fig:magerr_dens}. The solid lines are the
analytic fits to the data using the following function:
\begin{equation}
\mathrm{error}_i = c_1 \times \exp^{(c_2 \times \mathrm{mag} + c_3)} \, .
\label{eq:errorfunc}
\end{equation}
The values of the coefficients of the fits are given in Table~\ref{tab:errfits}.
The photometric measurements have large errors and show severe blending in region 1.
Therefore we will not consider this region further in the stellar populations analysis.
Region 2, which corresponds approximately to 2--4 kpc above the plane, and therefore is
strongly dominated by thick disk stars \citep{ibata+09}, has slightly larger errors, than
outer regions, but blending does not affect the $\sim 1-1.5$ mag below the RGB tip.

The analytic fits to the photometry errors derived from the 
completeness simulations for $450> \rho>140$ and 
$140>\rho >44$ fit well the errors derived from the overlap regions H1-H2 and H2-H3,
respectively, provided that the noise characteristics of the input images are properly taken
into account. This is a nice consistency check that the artificial star simulations
are providing a reliable estimate of photometric errors.

The completeness as a function of magnitude can be fairly well fitted using the following
analytical function \citep{fleming+95}:
\begin{equation}
f = \frac{1}{2} \left[ 1 - \frac{\alpha (m - m_0)}{\sqrt{1+\alpha^2(m-m_0)^2}} \right] \,.
\label{eq:complfitequ}
\end{equation}
The values of the fitted coefficients are given in Table~\ref{tab:complfits}.
We also derived the completeness relations as function of distance from the galaxy plane,
selecting stars only based on Z coordinates. The values of fitted coefficients are given
in  Table~\ref{tab:complfits} as well.
The plot and the tabulated values show that the region within $\sim 2$~kpc of the galaxy
plane is too crowded to yield accurate photometry even at the level of the red
giant branch tip. The area between 2 and 4~kpc from the plane has 50\%
incompleteness at an I-band magnitude of 27.27, while above 4~kpc above the disk
the radial dependence of incompleteness and magnitude errors is much weaker.

In spite of the relatively low Galactic latitude of the observed fields
($b=-17^\circ$), the contamination from the Milky
Way stars is negligible due to the small size of the ACS field. 
For example according to the Besan\c{c}on model of the Galaxy \citep{robin+03},
less than 170 Galactic stars are expected within the 3 ACS fields 
within the magnitude range $24<F814W<28.5$. 
This implies that the contamination from foreground
stars is $\la 0.05$\%. Most of the background 
sources are resolved due to the high resolution of the ACS images.


\section{Results}

In our analysis we use the distance  to NGC~891 of 9.73~Mpc \citep{mouhcine+07}, 
corresponding to a distance modulus of $(m-M)_0=29.94$~mag. 
The foreground reddening, obtained from  \citet{schlegel+98} maps is $E(B-V)=0.065$. 
This corresponds to $A_V=0.22$, and $A_I=0.13$~mag. 

In addition to the foreground extinction, the inner regions of NGC~891 suffer from 
dust extinction, which is readily visible in optical images covering the plane 
of the disk. Some amount of extinction is expected to be present also above the 
disk plane, given the extended distribution of the ionised gas, 
as well as the detected vertical distribution of molecular and HI gas 
\citep{kamphuis+07,scoville+93, oosterloo+07}. To correct for internal extinction 
within NGC~891, we proceed in the same way as in Paper~III: we compute the extra 
reddening $E(B-V)$ from the local value of the HI column density using the high 
resolution deep HI map of \citet{oosterloo+07} and assuming the conversion 
between HI column density and $E(B-V)$ derived for the MW by \citet{rachford+09}. 
Then, half of this additional $E(B-V)$ is added to the foreground MW reddening 
value. In assigning only half of the extra-extinction derived from the HI map we 
assume that approximately half of the gas (and dust) is in front, and half behind 
the observed stars in NGC~891. 


\begin{figure*}
\resizebox{\hsize}{!}{
\includegraphics[angle=0]{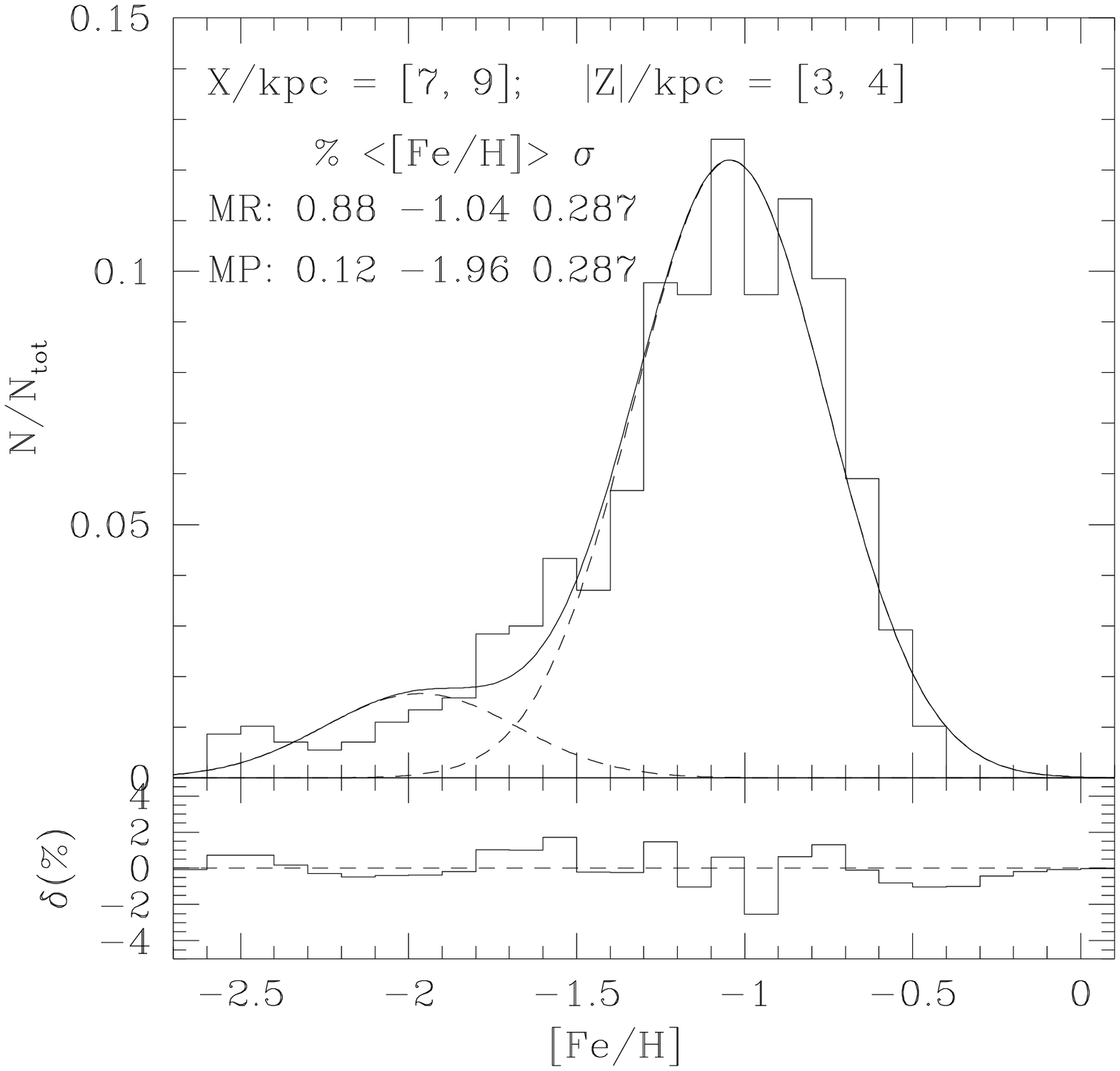}
\includegraphics[angle=0]{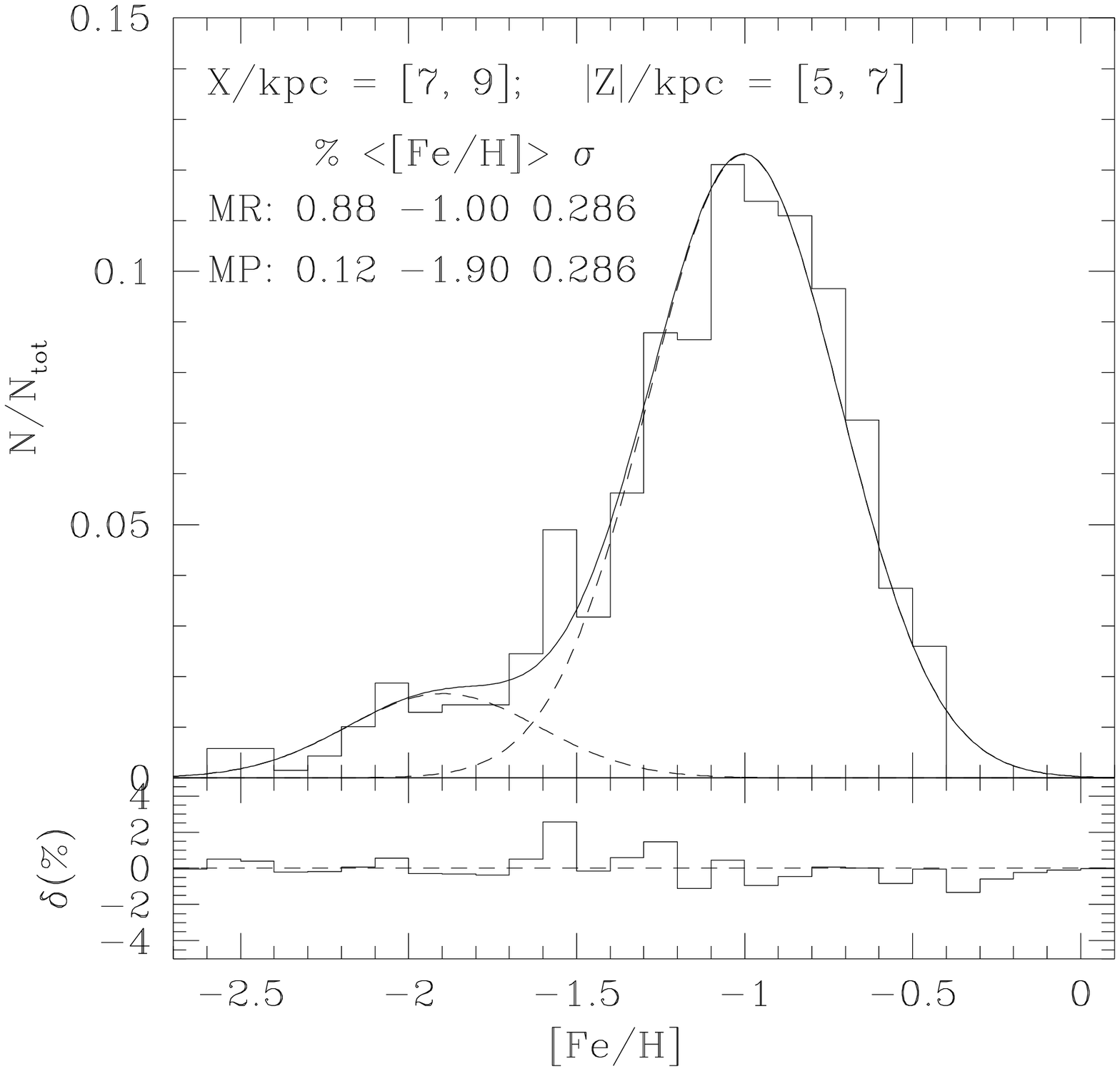}}
\caption{Normalised, extinction corrected, MDFs of stars at solar cylinder-like 
distances at two different heights from the disk of NGC 891, 3-4~kpc (left panel) 
and 5-7~kpc (right panel). Solid lines show the best double Gaussian function fits 
based on the KMM statistical test. Indicated are the best fit parameters of each 
of the two components of the fit, referred to as the metal-poor (MP) and the 
metal-rich (MR), and shown as dashed lines. }
\label{fig:mdf_x7_9}
\end{figure*}

The impact of the absorption correction uncertainty is minimised in the
following by investigating primarily the areas beyond 2 kpc above the
disk plane, where most of the dust and gas is located. The derived
metallicity and colour gradients are consistent within the $2 \sigma$
error-bars if the additional extinction contribution is varied by about
a factor of two.  If less extinction is assumed the gradient is steeper,
while it flattens if extinction is higher than assumed. This is primarily
due to the change in colour/metallicity in the inner 2.5-3 kpc.

After applying this extinction correction to our photometry, we investigate first 
the stellar population content at the distance corresponding to the solar circle. 
After that we discuss the metallicity distributions of the thick disk, the inner 
spheroid and the halo of NGC~891.  To minimise the photometric errors and crowding 
we limit our analysis to stars located in regions with stellar density lower than 
650 $N(RGB)/\mathrm{kpc^2}$.

\subsection{Metallicity distribution function}

Both stellar evolution models and globular cluster observations show that colours 
of red giant stars are much more sensitive to metallicity than age, provide thus 
an excellent way to estimate the metallicity distribution of a stellar population.
This approach has been widely used in the literature 
\citep[e.g.][]{harris+99,saviane+00_rgb,zoccali+03,mouhcine+05c,rejkuba+05}, and 
is calibrated using Galactic globular clusters, and stellar evolutionary tracks.
Its drawback is that one cannot separate stars belonging to the early-AGB (E-AGB) 
evolutionary phase, which are located along the RGB, with colours bluer with 
respect to  the first ascent giants of the same age. 
In a composite stellar population, with a range of metallicities, the E-AGB stars 
of a metal-rich population may overlap in colour with metal-poor RGB stars. 
However, the life-time of AGB stars is significantly shorter, and one expects only one 
E-AGB star for 40 RGB stars in an old population \citep[e.g.][]{renzini98}. 
Moreover, at the blue edge, E-AGB stars belonging to the metal-poor population 
are excluded because they are bluer than the most-metal poor evolutionary track. 
An additional complication is due to a possible mix of ages in our fields. 
In the interpolation, we assume a single old age for all the stars, which may 
introduce a bias of up to 0.1 dex, if the age is 4 Gyr younger \citep{rejkuba+05}.  
From comparison with simulated CMDs, and from the absence of blue plume stars  
and a large fraction of thermally-pulsing AGB (TP-AGB) stars, the fractions of 
stars younger than $\sim 6$~Gyr are expected to be small in both the thick disk
and the halo of NGC~891.

Metallicities for each star were derived by interpolating between the set of 
$\alpha$-enhanced RGB tracks for stars with $0.8$~M$_\odot$ \citep{vandenberg+06}. 
For more details we refer to \citet{mouhcine+05c,mouhcine+07}. We note here that 
only stars with colours corresponding to the metallicity range between 
$\mathrm{[Fe/H]}=-2.314$ and  $-0.397$, that satisfy our selection criteria, 
and have $I$-band magnitudes brighter than 26.5~mag (to avoid as much as possible 
completeness corrections), have been used to construct the metallicity 
distribution functions (MDFs). From the comparison of the observed CMDs with the 
stellar evolutionary isochrones it is clear that there are almost no stars with 
metallicity above Z=0.008 isochrone (see below), which corresponds to 
$\mathrm{[Fe/H]}=-0.38$. Therefore we expect that this choice of tracks used for 
interpolation, and the fact that we do not extrapolate beyond the validity of the 
models, does not affect the metal-rich end of our MDF.


\begin{figure*}
\resizebox{\hsize}{!}{
\includegraphics[angle=0]{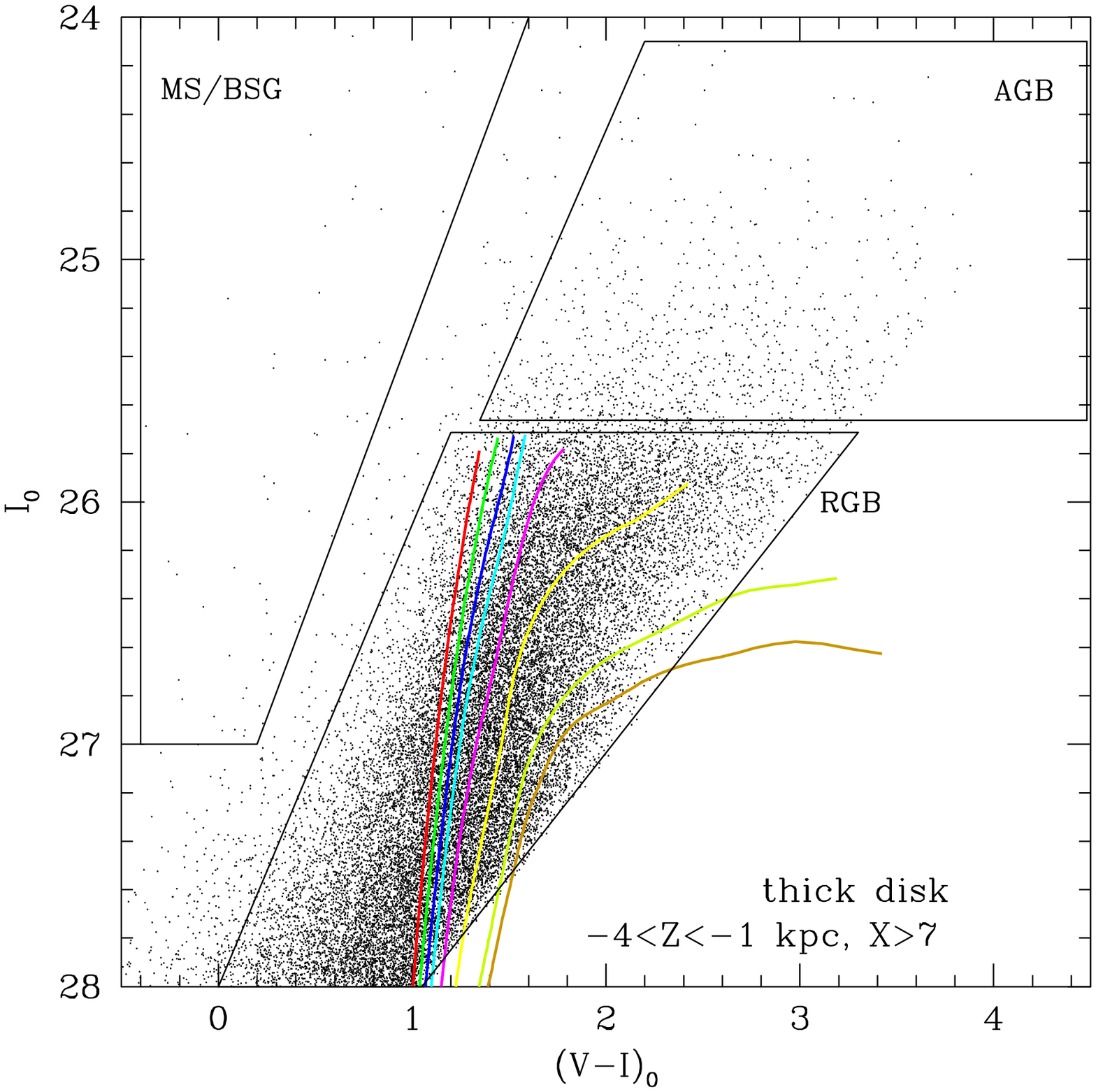}
\includegraphics[angle=0]{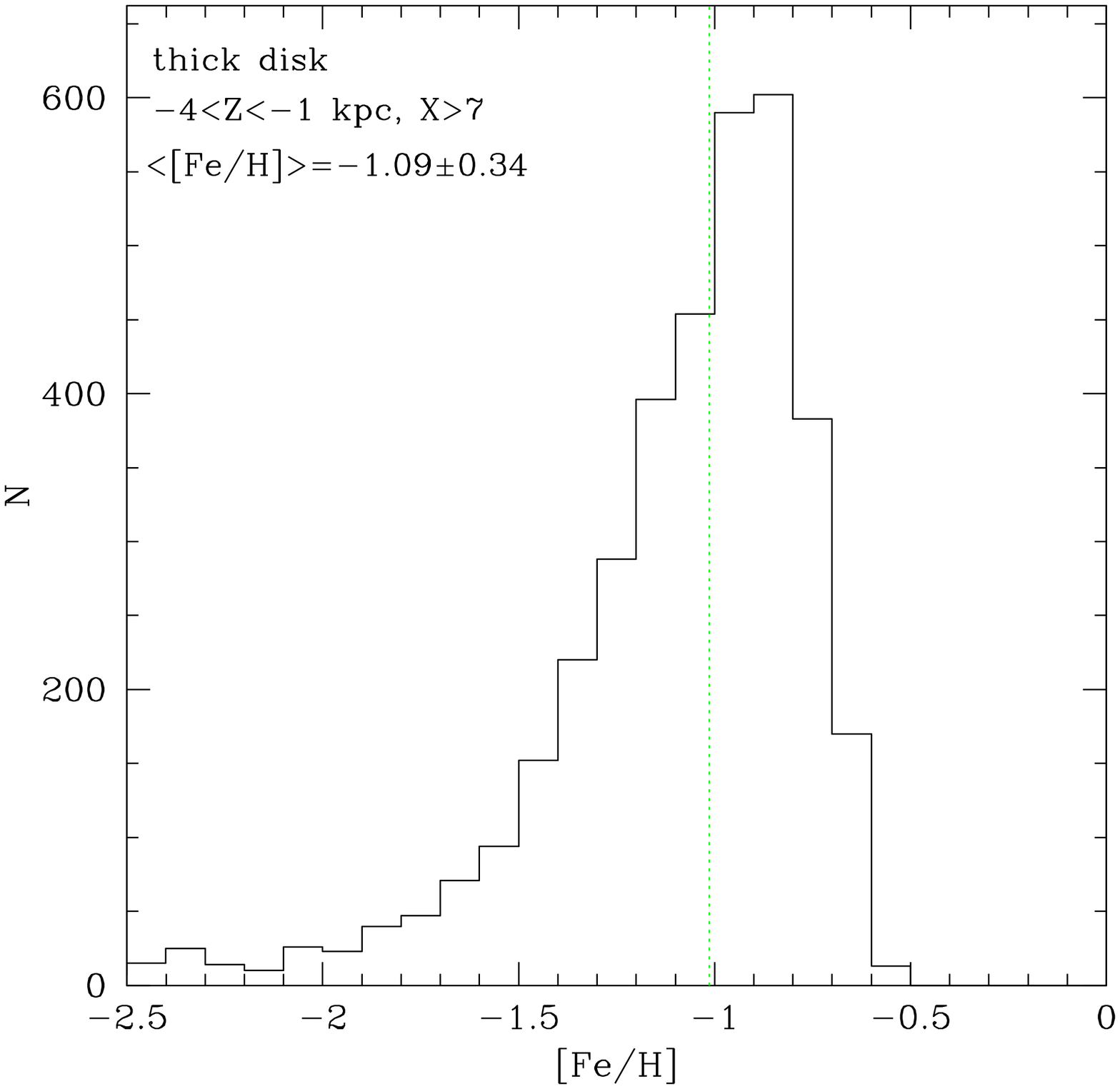}}
\caption{Left: Colour-magnitude diagram for the selected thick disk stars in 
NGC 891. Overplotted are BASTI solar scaled isochrones \citep{pietrinferni+04}
for a 10 Gyr old population with metallcities $Z=0.0001, 0.0003, 0.0006, 0.001,
0.002, 0.004, 0.008, 0.01$.
Right: MDF of thick disk stars. Dotted vertical line indicates median [Fe/H] 
value, while the average [Fe/H] and $1\sigma$ dispersion around the mean is 
shown in the upper left corner.}
\label{fig:cmd_mdf_thick}
\end{figure*}

\subsection{Solar-cylinder-like populations}

Figure~\ref{fig:mdf_x7_9} shows the MDFs of stars selected at $8\pm1$~kpc distance 
along the major axis and at two different heights above the plane of the galaxy: 
at $3-4$~kpc (left panel) and $5-7$~kpc (right panel). The histograms are normalised, 
extinction-corrected stellar metallicity distribution functions in the (X,Z) regions 
indicated in each panel. Both distributions have a prominent relatively metal-rich 
peak with a median of [Fe/H]$\sim -1.0$~dex and a sparsely populated metal-poor tail. 
The properties of metallicity distributions are quantified using the widely used 
Kaye's Mixture Model (KMM) statistical test \citep[][and references therein]{ashman+94}. 
The KMM test uses the maximum likelihood technique to test if a distribution is better 
modelled as a sum of two Gaussian than as a single Gaussian (the null hypothesis). 
\citet[][]{ashman+94} have cautioned that the output likelihood in the case where 
the two populations have different dispersions is difficult to interpret, so we have 
assumed that both populations have similar dispersions. The solid line shows 
the best double Gaussian distribution fit, which is preferred to a single Gaussian 
distribution based on the KMM statistics. Indicated are the parameters of the best 
fit models as assigned by the KMM test for the two components, referred to as the 
metal-poor and the metal-rich component and shown by the dashed lines. 


\begin{figure*}
\resizebox{\hsize}{!}{
\includegraphics[angle=0]{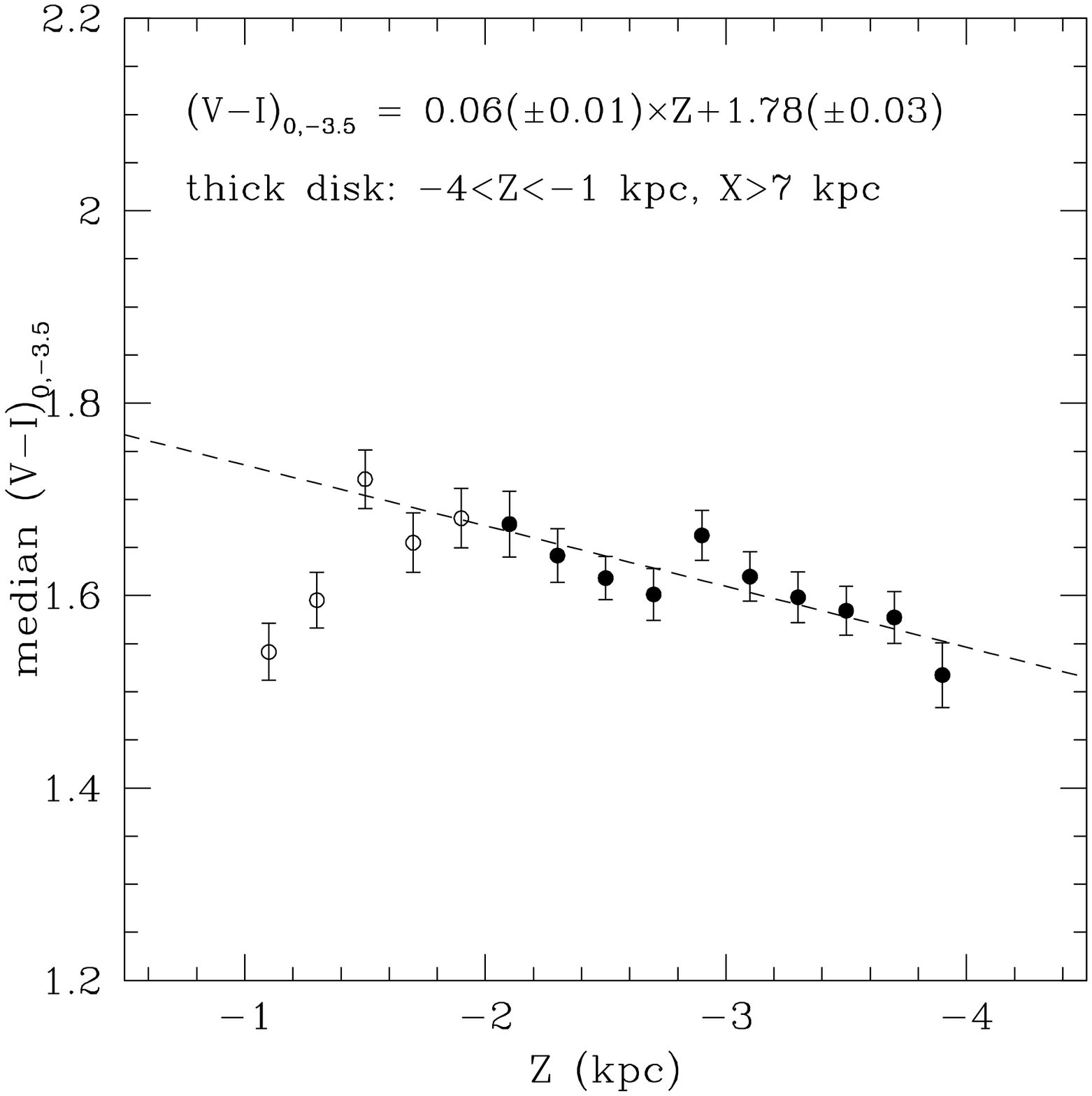}
\includegraphics[angle=0]{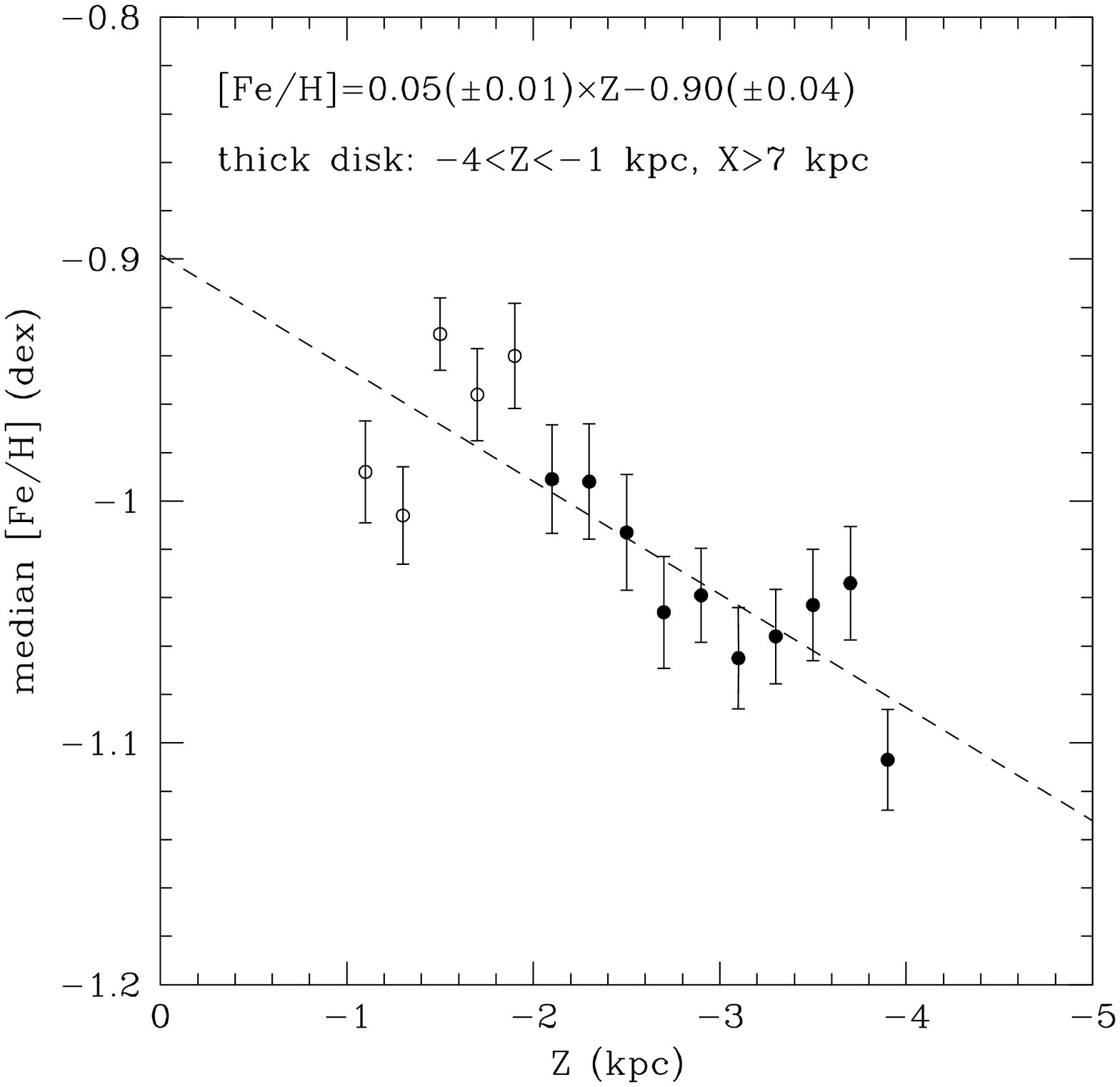}}
\caption{Colour gradient (left panel) and median metallicity gradient (right panel) 
of the thick disk stellar population ($-4<Z<-1$~kpc and $|X|>7$~kpc) perpendicular 
to the galactic plane. The dashed line in each panel shows the linear fit to the 
data, restricted to the solid points. }
\label{fig:zgrad_thick}
\end{figure*}

The distribution functions shown in Figure~\ref{fig:mdf_x7_9} are significantly 
different from those of the Milky Way stars at similar locations. 
\citet{ivezic+08} have shown for stars with $7\,{\rm kpc} < R < 9\,{\rm kpc}$, 
that the metal-rich components, with ${\rm <[Fe/H]>\sim -0.7}$, dominate only 
at relatively small vertical distances from the plane, i.e.,$|Z|<2\,{\rm kpc}$. 
For higher vertical distances, i.e., $|Z|\ga 5\,{\rm kpc}$ the fraction of stars 
belonging to the metal-rich component is vanishingly small, and are completely
out-numbered by stars with ${\rm <[Fe/H]>\sim -1.5}$. NGC~891 stars at similar 
distances along the minor and the major axes exhibit a different behaviour. 
The properties of the metal-rich and the metal-poor components of the stellar MDF 
do not change significantly as a function of the vertical distance from the galactic
plane. Metal-rich stars are still present in abundance at a vertical distance beyond 
5~kpc, in stark contrast with the stellar content of the Galaxy at similar distances.
 
The lower panels of Figure~\ref{fig:mdf_x7_9} show the residuals from the double
Gaussian fits of the observed metallicity distributions. No systematic deviations 
are noticeable. The difference between the observed MDF of the MW stars at the 
solar cyllinder and the double Gaussian fit shows the presence of an additional 
population of intermediate metallicity stars, i.e., [Fe/H]$\sim -1.0$~dex, a 
reminiscent of the so-called metal-weak thick disk \citep[][]{morrison+90}. 
The metal-rich peak of the MDF of the MW stars within the solar cyllinder is best 
fitted by a double Gaussian model up to $|Z|\sim 4\,{\rm kpc}$. For NGC~891 however, 
the metal-rich peak of stars at the solar cyllinder-like locations is satisfactorily
modeled as a single component, indicating the absence of a distinct stellar 
population with intermediate metallicities within $|Z|\sim\,3-6$\, kpc.

The observed differences could be due to that thick disk stars, which should 
be the dominant contributors at $Z=3-4$~kpc, are still present beyond 5~kpc 
(see Fig.~7b of Paper~III). In addition, the stellar halo of NGC~891 has a higher 
mean metallicity than that of the MW at comparable radial and vertical distances 
(Paper~I and also Sect.~\ref{sect:spheroid} below). All this points toward the 
presence of a mix of populations that could be due to more massive accretion 
events than is typical in the Galaxy at the solar neighbourhood. While this may sound like 
a far-fetched conclusion, the occurrence of recent accretion events is supported 
by the detection of significant small scale substructures across the thick disk 
and spheroid of NGC~891 (Paper~III).

\subsection{Thick disk stellar population} 
\label{sect:thickpops}

The structural analysis presented in Paper III has shown that a thick disk component 
is present in NGC~891, with stellar densities well fitted by exponential profiles 
both vertically, with a scale height of $h_Z=1.44 \pm 0.03$~kpc, and radially, with 
a scale length of $h_R=4.8\pm0.1$~kpc. The presence of an inner stellar spheroid, 
combined with large stellar densities, causing large photometric uncertainties, 
incompleteness, and significant blending (density region 1, with $\rho>1200$~stars 
extends almost up to 4~kpc above the plane along the minor axis), prevent the 
detection of the thick disk population within the inner $\sim5-6$ kpc. To investigate 
the properties of the thick disk stellar content, as free as possible from the bulge 
and halo stars, we select stars with $-4 < Z < -1$~kpc, and $|X|> 7$~kpc 
\citep[in particular see their Fig.~4 and 7]{ibata+09}. As already mentioned, to 
minimize the photometric errors and crowding we limit our analysis to stars located 
in regions with stellar density lower than 650 $N(RGB)/\mathrm{kpc^2}$. 


\begin{figure}
\resizebox{\hsize}{!}{
\includegraphics[angle=0]{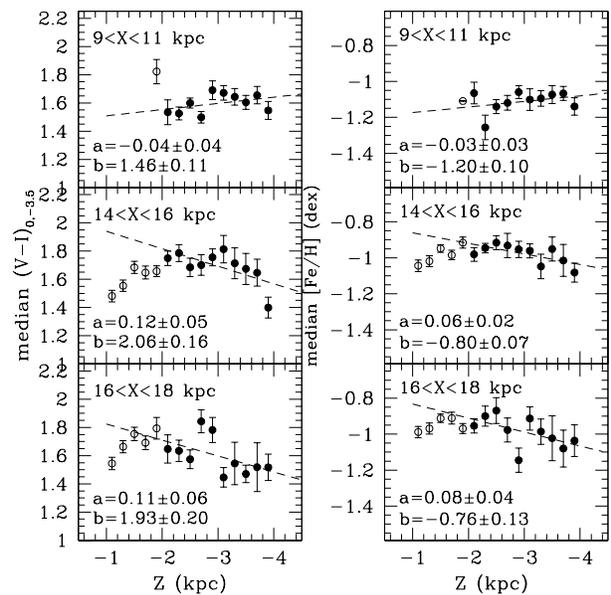}}
\caption{Colour gradient (left panels) and median metallicity gradient (right panels) 
of the thick disk stars at three different locations along the major axis  perpendicular to the galactic plane
as indicated in each panel.  The $a$ and $b$ values indicated 
in each panel are the coefficients of the linear least squares fit. }
\label{fig:plotzgrad_reg_thick}
\end{figure}


\begin{figure*}
\resizebox{\hsize}{!}{
\includegraphics[angle=0]{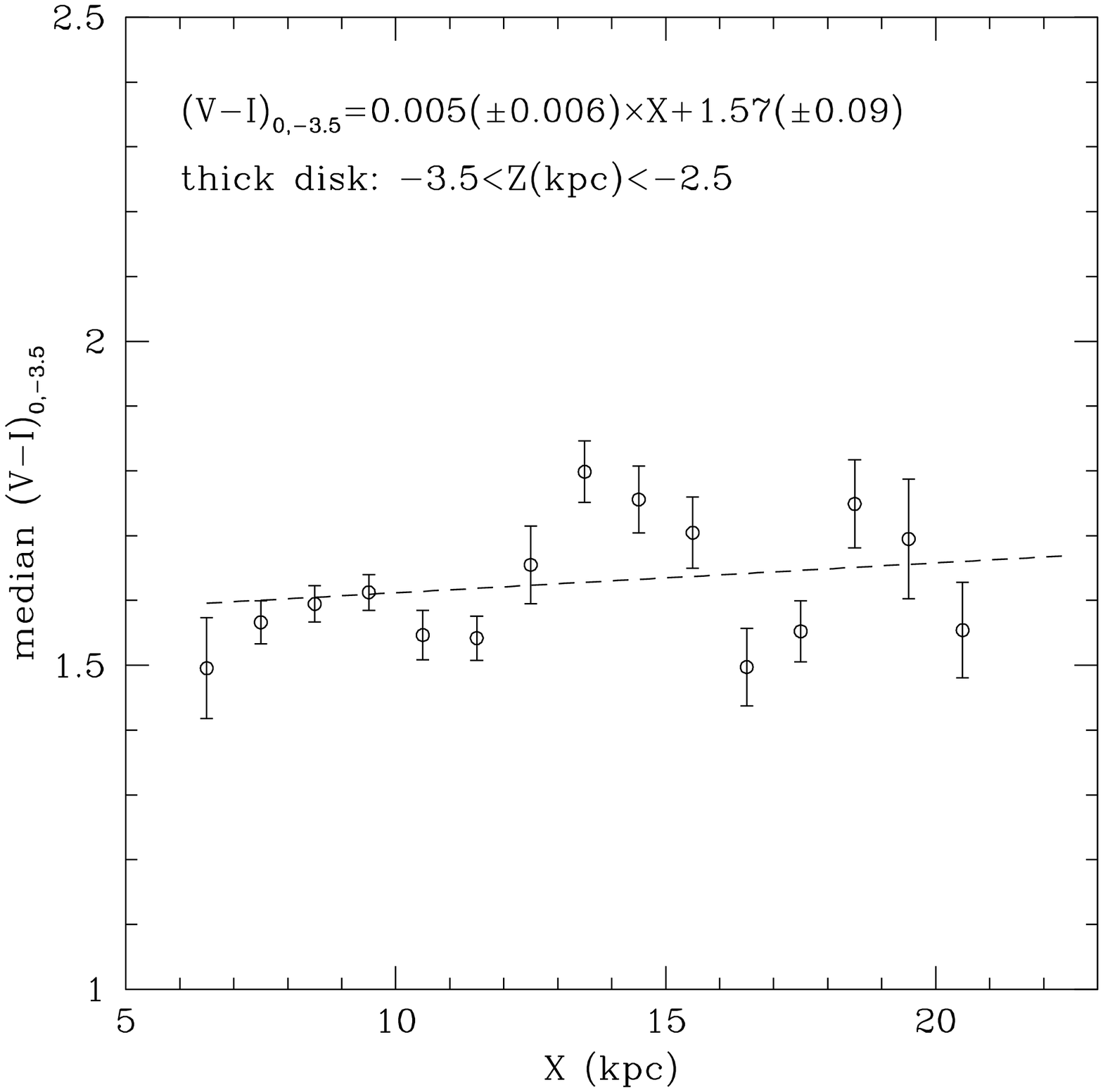}
\includegraphics[angle=0]{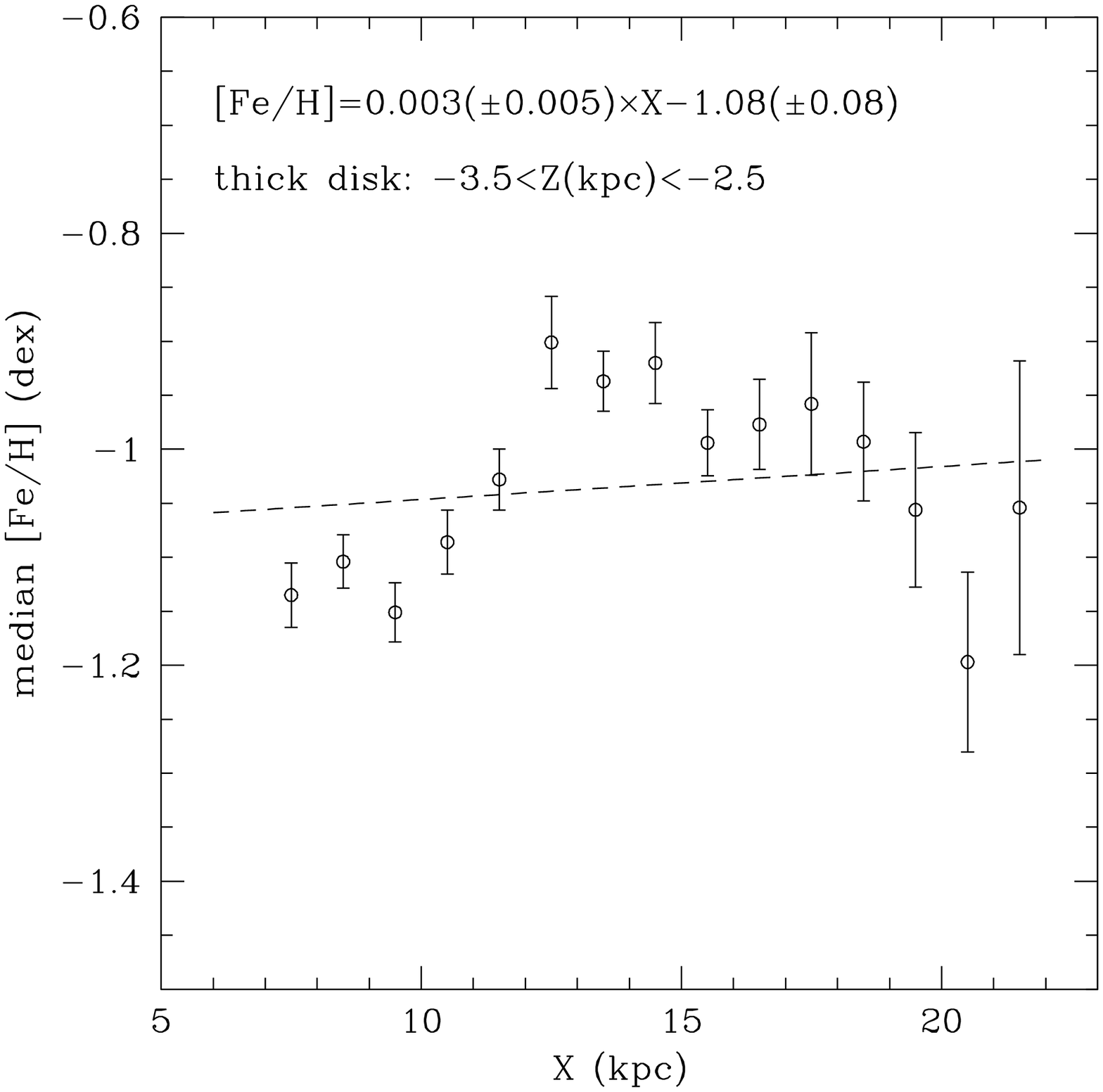}}
\caption{Median colour (left panel) and median metallicity (right panel) of the thick 
disk stars as a function of the major axis distance. Only the thick disk stars between 
2.5 and 3.5 kpc above the plane are selected in this plot.}
\label{fig:xgrad_thick}
\end{figure*}

\input{ngc891_paper2_tab05}


\begin{figure*}
\resizebox{\hsize}{!}{
\includegraphics[angle=270]{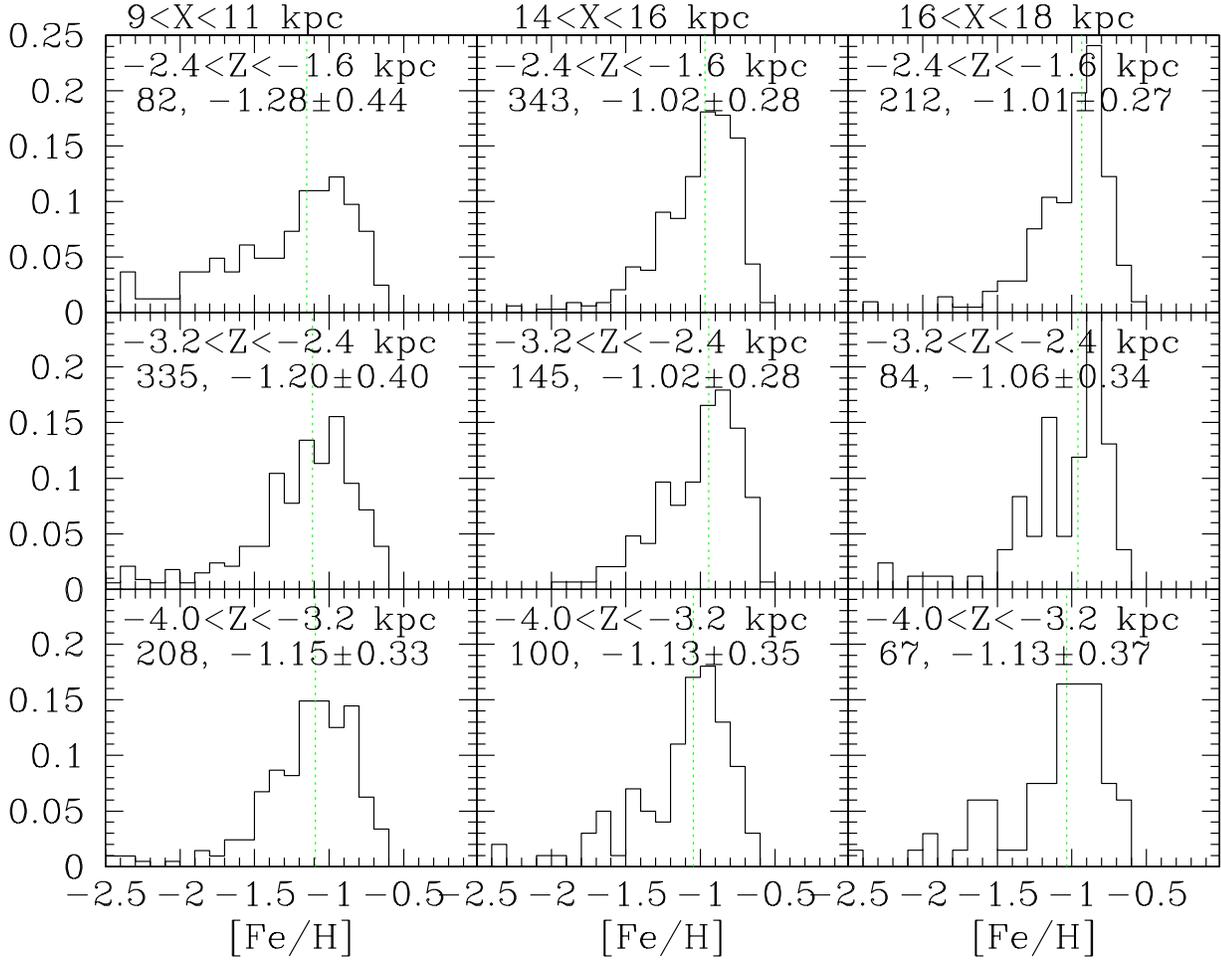}}
\caption{Normalised metallicity distribution functions of thick disk RGB 
stars selected to lie at different heights above the plane (upper row: $-2.4<Z<-1.6$ kpc; 
middle row:  $-3.2<Z<-2.4$ kpc; and bottom row: $-4.0<Z<-3.2$~kpc), and at different 
distances along the major axis (left column: $9<X<11$~kpc; middle column: $14<X<16$~kpc; 
and right column: $16<X<18$~kpc). The mean [Fe/H] and 
$1\sigma$ dispersion around the mean are indicated in each panel together with the 
number of stars in each region. The vertical dotted lines indicate the median [Fe/H].}
\label{fig:mdf_distributionsXZ_thick}
\end{figure*}


\begin{figure*}
\resizebox{\hsize}{!}{
\includegraphics[angle=0]{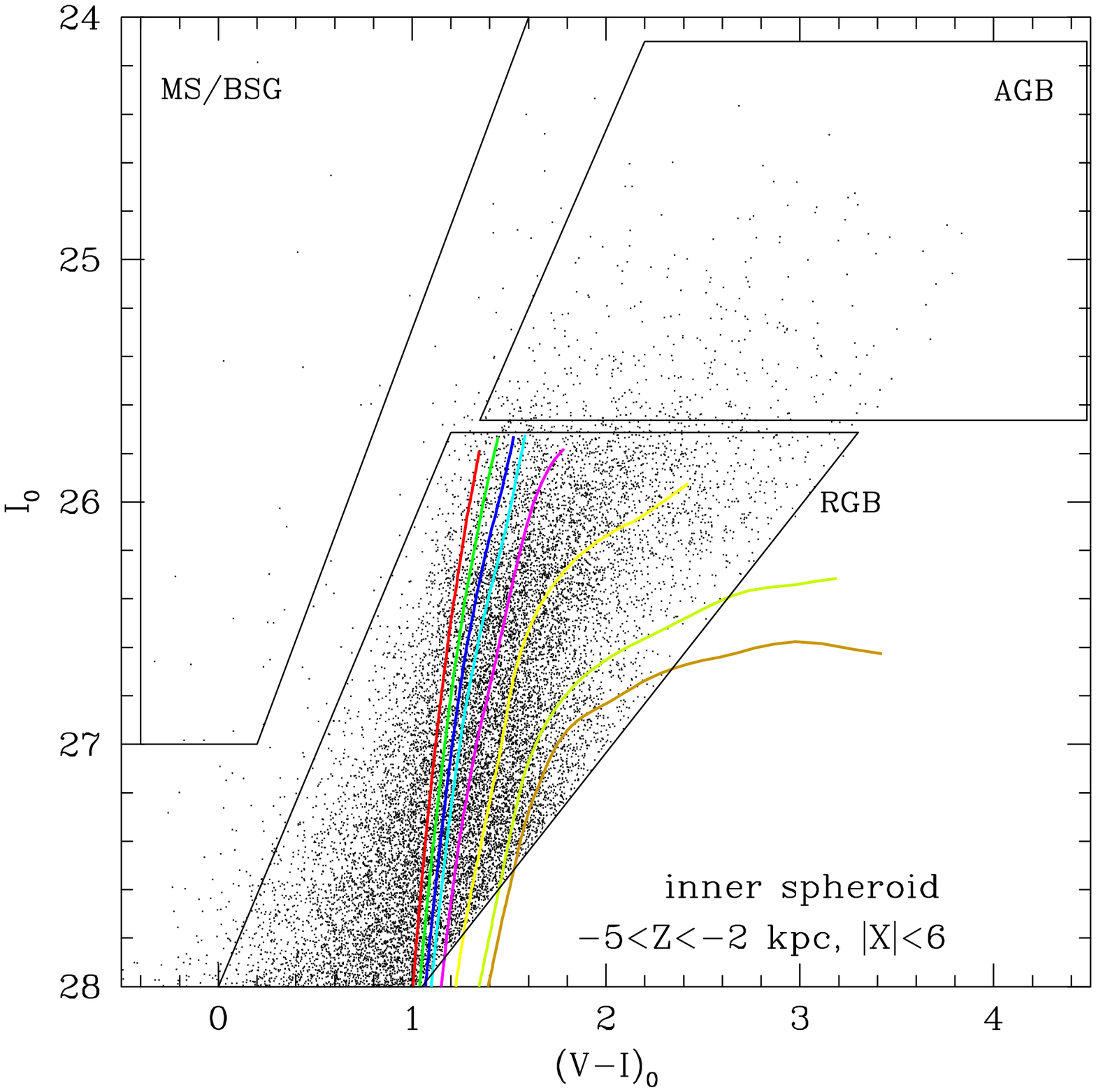}
\includegraphics[angle=0]{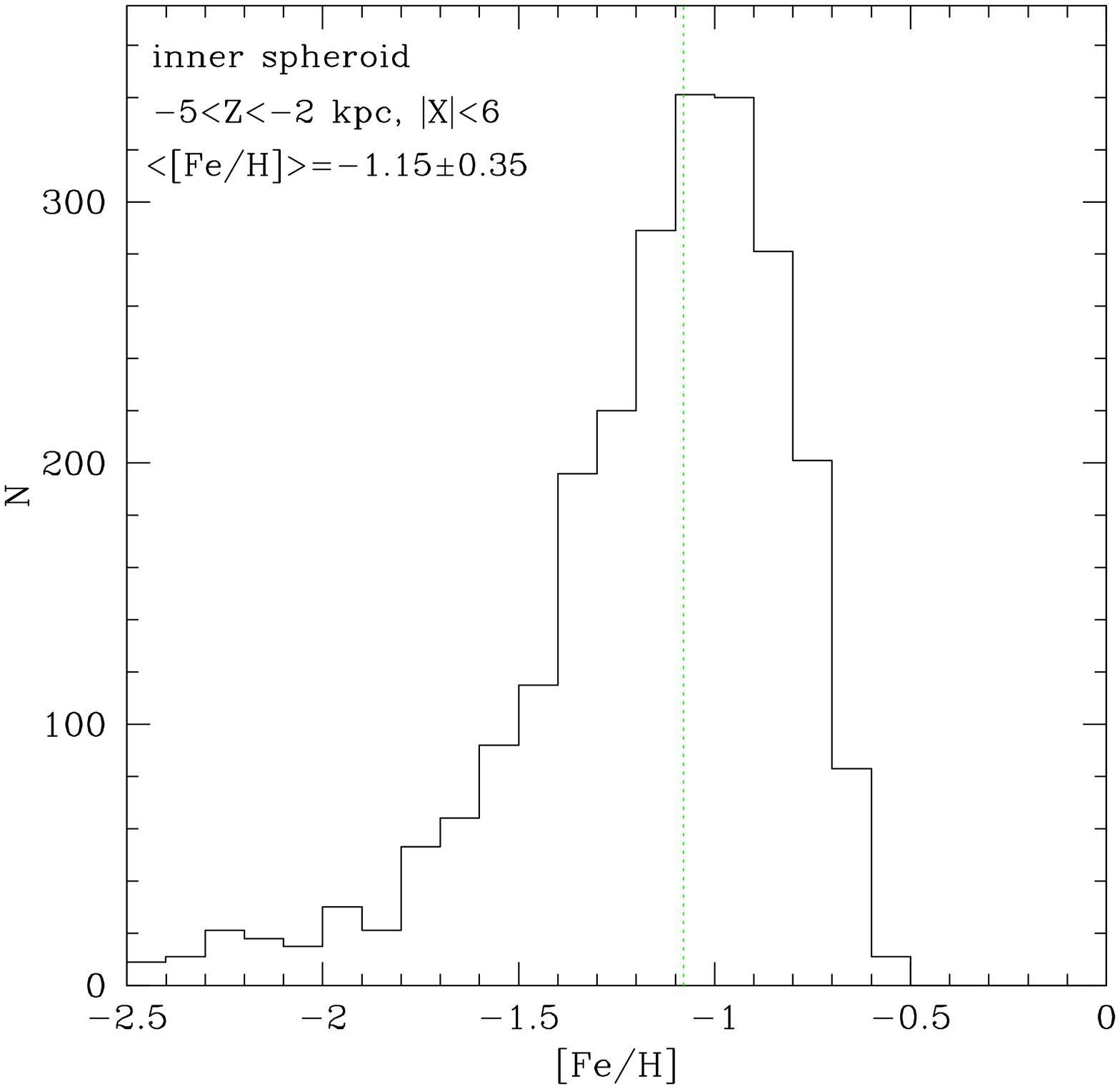}}
\caption{Similar to Fig.\,\ref{fig:cmd_mdf_thick} but for the inner spheroid
stellar populations.}
\label{fig:cmd_mdf_innersph}
\end{figure*}

The colour-magnitude diagram of selected thick disk stars is shown in 
Fig.~\ref{fig:cmd_mdf_thick}. The most prominent feature is the well-populated red 
giant branch, with stars covering a wide range of colours. In the area of the 
diagram where blue plume stars, belonging to the young main sequence, and blue 
super-giant stars are expected to be located, there are only a handful of stars, 
which could be either scattered there by the combination of photometric errors and 
blending, or could be mis-identified background or foreground sources. Overplotted on 
the CMD are Teramo stellar evolutionary isochrones for a 10 Gyr old stellar population 
with the range of metallicities 
$Z=0.0001, 0.0003, 0.0006, 0.001, 0.002, 0.004, 0.008, 0.01$. While the  incompleteness 
prevents the detection of stars with metallicities higher than half-solar, only very few, if 
any, are expected, given that the density of stars decreases strongly between the $Z=0.004$ 
and $Z=0.008$ isochrones. The right panel of Fig.~\ref{fig:cmd_mdf_thick} shows the 
MDF of all thick disk stars with $I$-band magnitudes brighter than 26.5~mag. The peak 
of the stellar metallicity distribution of the thick disk stars is around $-0.9$~dex, 
with a mean metallicity of $-1.09$~dex and one $\sigma$ scatter of $0.34$~dex. 
The median of the stellar metallicities 
of the thick disk is $\mathrm{[Fe/H]}=-1.01$~dex, slightly lower than that measured 
for the MW thick disk of $-0.8$~dex for stars at comparable vertical distances above 
the galaxy plane \citep{gilmore+95,ivezic+08}. It worth mentioning that thick disk 
stars selected here are distributed over a wide range of radial distances, i.e., 
${\rm 7\la X/kpc \la 22}$, while the samples of the Galaxy thick disk stars are 
restricted generally to the solar neighbourhood. 

The spatial variation of the properties of thick disk stars holds important clues
on the formation mechanism(s) of this disk component. The size of the sample of thick 
disk stars selected here is large enough to permit a study of the vertical and radial 
variation of their properties. The left panel of Fig. \ref{fig:zgrad_thick} shows the
variation of the median colour of thick disk stars with $M_I=-3.5\pm 0.1$~mag, i.e., 
$(V-I)_{0,-3.5}$ a powerful metallicity indicator for old stellar populations, along
the vertical direction. The dashed line shows a linear fit to the data where only 
solid dots ($Z\leq 2$~kpc) are used in the fit. Despite our efforts to correct for incompleteness 
due to crowding and large extinction in the inner parts, we have chosen to exclude
stars in those regions for the sake of obtaining clear conclusions. The right panel of 
Fig. \ref{fig:zgrad_thick} shows the variation of the median metallicity along the 
vertical direction. A strong vertical gradient of the median colour and 
metallicity of thick disk stars is present. The detected vertical gradient could 
be genuine, or alternatively, given the wide range of radial distances covered by 
stars in the thick disk sample, could be due to a mix of different populations with 
radially varying contributions. The inspection of the vertical stellar density 
profiles indicates that the contribution of halo stars to the overall populations 
with $|Z|<4\,{\rm kpc}$, estimated from the extrapolation of the star count profiles 
stars with $|Z|>6\,{\rm kpc}$, changes as one moves radially away from the minor 
axis (see Fig. 7 and Fig. 8 of Paper~III). To investigate this further, we plot the 
evolution of both the median colour $(V-I)_{0,-3.5}$, and median metallicity of thick 
disk stars as a function of the vertical distance in three radial bins in 
Fig.~\ref{fig:plotzgrad_reg_thick}: between 
$9-11$~kpc (upper panel), $14-16$~kpc (central panel) and between $16-18$~kpc 
(bottom panel). In each panel we indicate the slope and the normalisation of the 
linear fit to the data, restricted to the solid dots. The results of this exercise 
are given in a tabular format in Table~\ref{tab:vimdf_thick} reporting the mean, 
median and rms dispersion of the $(V-I)_{0,-3.5}$ colour and [Fe/H] distributions 
at a range of distances along the minor axis (Z range) and major axis (X range). 
The thick disk stars with $X\la11~$kpc, where the contamination from halo stars 
is expected to be small, do not show vertical variations of their median metallicities 
and colours. This is similar to what is observed for the Galactic thick disk stars 
at similar radial distances and heights from the galactic plane \citep{ivezic+08}. 
For thick disk stars well away from the minor axis, i.e., $X\ga14~$kpc, marginally 
significant mild vertical gradients are present however. The outer spheroid stellar 
populations shift the average colour of stars close to the edge of the thick disk 
outside 14~kpc towards bluer values, lowering therefore the median metallicities.

Fig.\,\ref{fig:xgrad_thick} shows the variation of the median colour (left panel) 
and median metallicity (right panel) of thick disk stars along the radial direction, 
for those with ${\rm -3.5<Z/kpc<-2.5}$. The dashed lines show the linear fits to the 
data. No radial metallicity gradient is detected for thick disk stars. However, the 
median metallicities/colours are found to vary significantly within the narrow bin
of vertical distances. The mean thick disk colour for the vertically selected stars 
is $(V-I)_{0,-3.5}=1.65$, with $1\sigma$ scatter around the mean of $0.13$~mag. 
The dispersion around the mean metallicity is approximately $0.35$~dex, and does 
not vary much as one moves along the major axis.

\begin{figure*}
\resizebox{\hsize}{!}{
\includegraphics[angle=0]{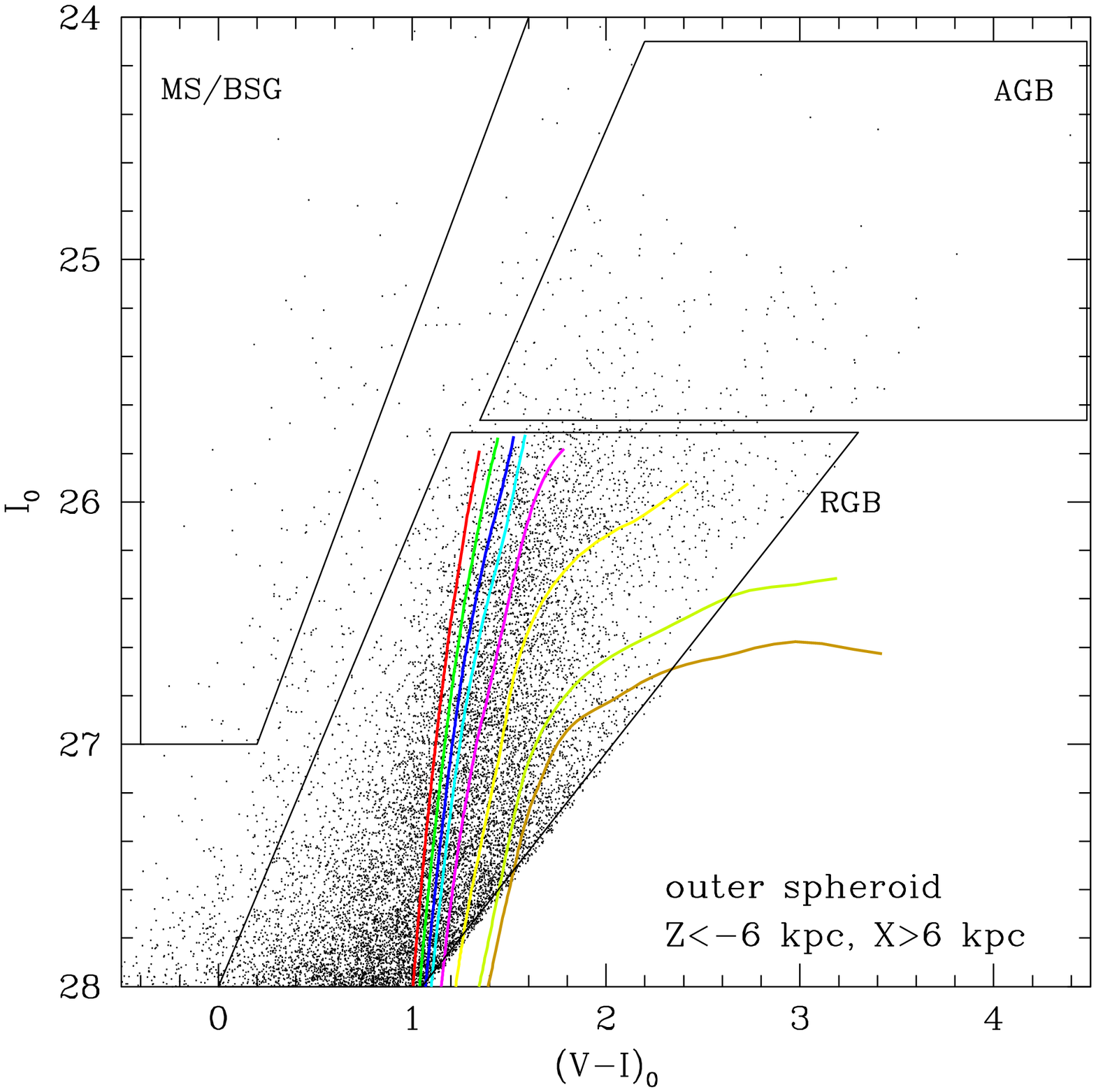}
\includegraphics[angle=0]{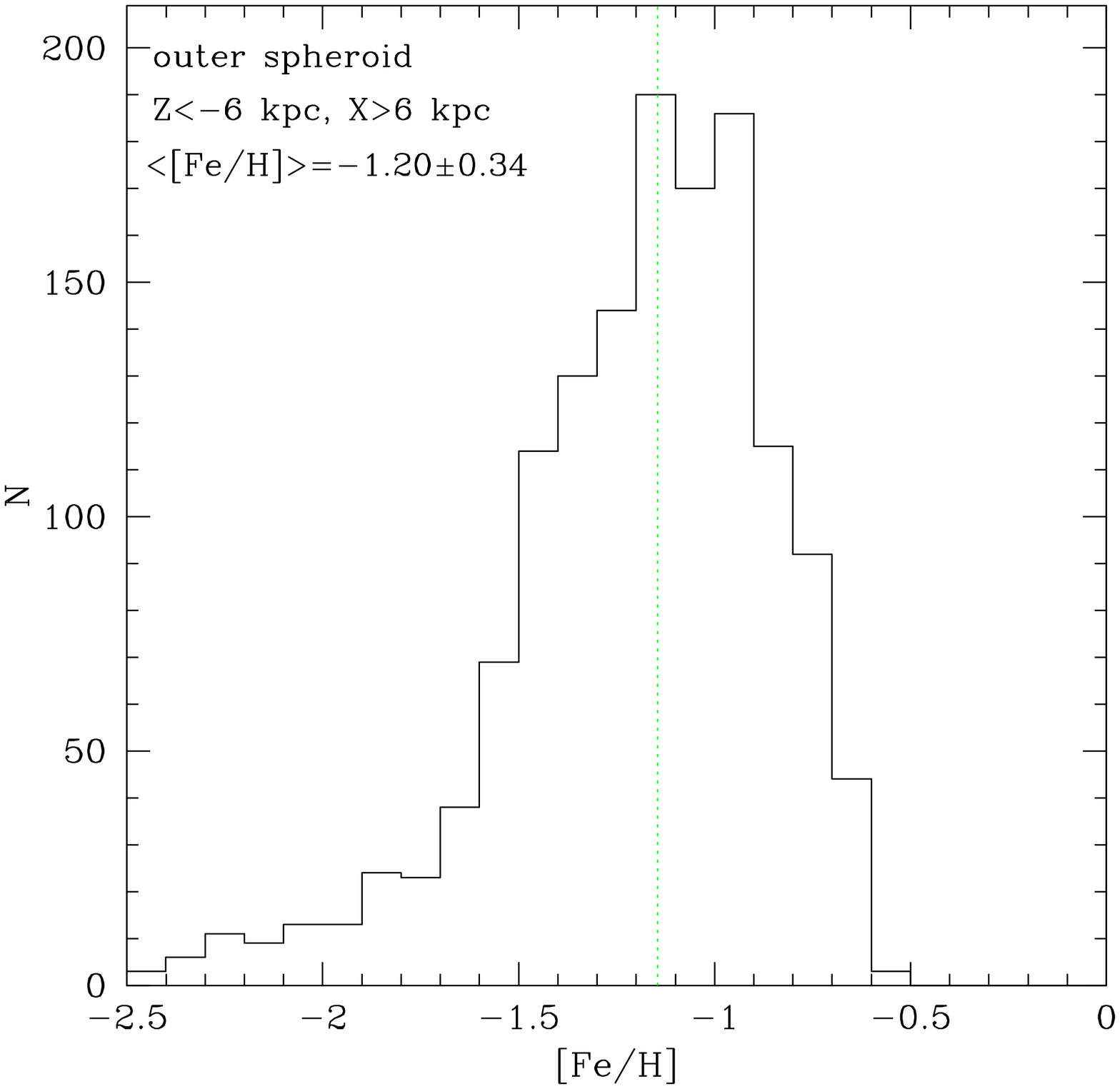}}
\caption{Similar to Fig.\,\ref{fig:cmd_mdf_thick} but for the halo stellar 
populations.}
\label{fig:cmd_mdf_halo}
\end{figure*}


\begin{figure*}
\resizebox{\hsize}{!}{
\includegraphics[angle=0]{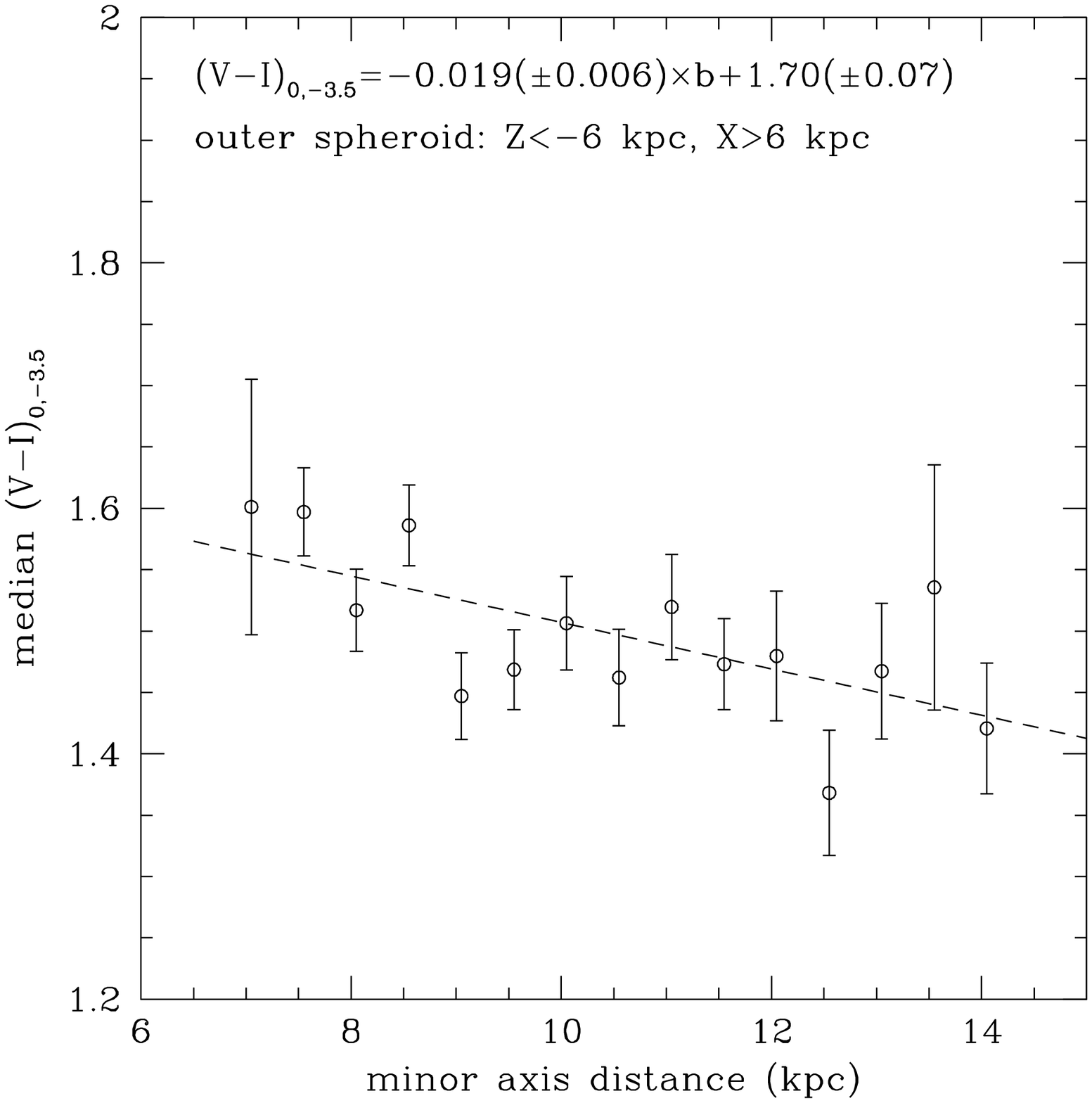}
\includegraphics[angle=0]{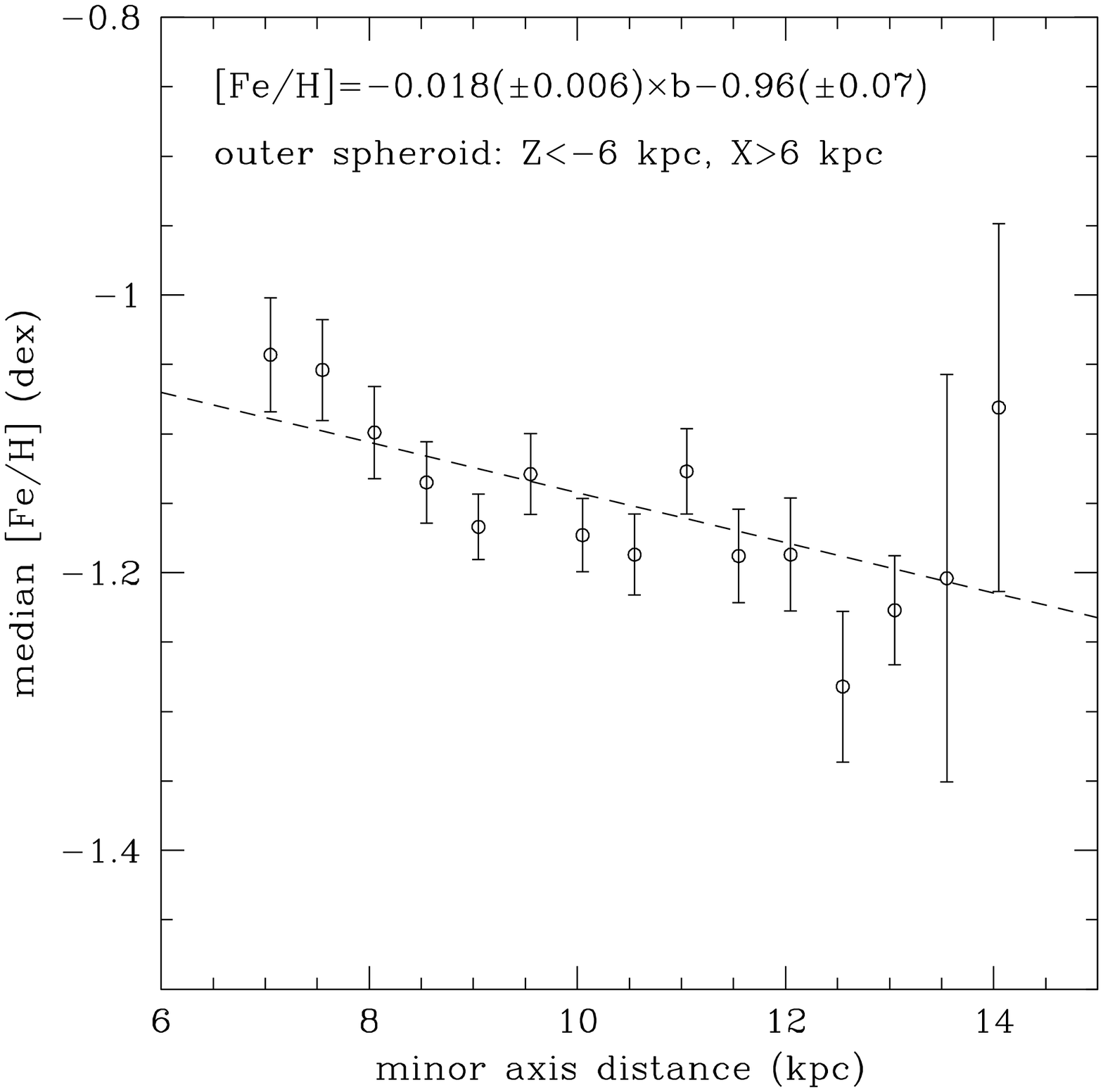}}
\caption{Median colour (left panel) and median metallicity (right panel) of the outer 
spheroid stars computed in elliptical rings with axis ratio 0.5 are plotted as a 
function of minor axis distance (b) from the centre of NGC~891. Error-bars indicate 
errors on the mean.}
\label{fig:mdfgrad_halo}
\end{figure*}

Figure~\ref{fig:mdf_distributionsXZ_thick} shows the normalised, extinction corrected, 
MDF histograms for thick disk RGB stars, and selected to lie at radial distances
$9<X<11$~kpc (left column), $14<X<16$~kpc (central column), and $16<X<18$~kpc (right 
column). The thick disk stars at different radial distance bins are split into 
different vertical heights $-2.4<Z<-1.6$~kpc (top row), $-3.2<Z<-2.4$~kpc (central 
row), and $-4.0<Z<-3.2$~kpc (bottom row). The figure shows that the properties of 
the metallicity distribution, i.e., the metallicities of the metal-poor and the 
metal-rich peaks and their relative contributions, of thick disk stars do not appear 
to change either radially or vertically.

\subsection{The stellar population of the spheroid}
\label{sect:spheroid}


\begin{figure*}
\resizebox{\hsize}{!}{
\includegraphics[angle=270]{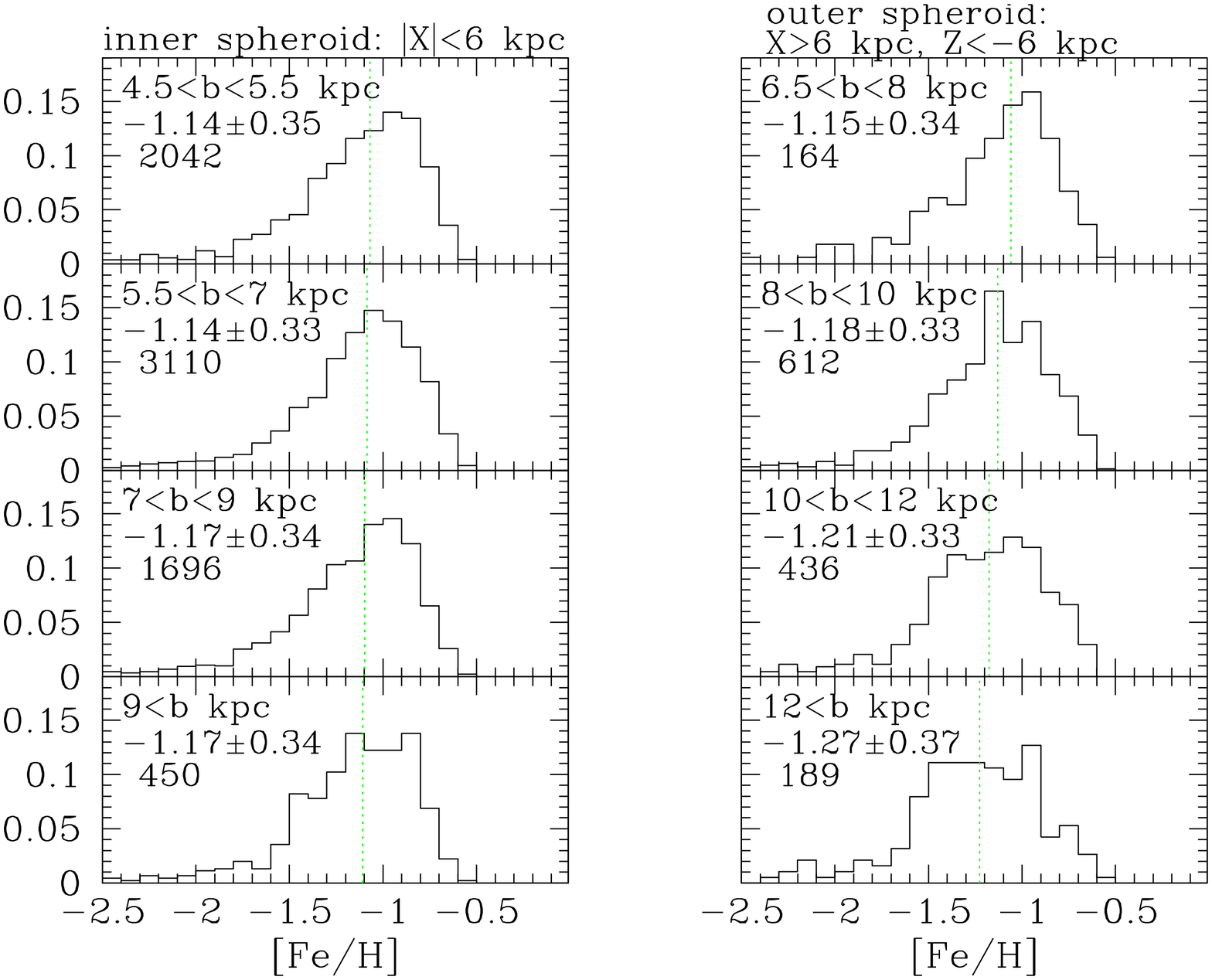}}
\caption{Normalised metallicity distribution functions of the inner spheroid stars are 
plotted in the panels on the left, and the MDFs of the outer spheroid stars are plotted 
in the panels on the right. In each panel stars are selected in elliptical annuli with 
limits according to the minor axis size indicated. For the inner spheroid axis ratio 
of the ellipses is 0.73, and for the outer spheroid it is 0.5. The mean [Fe/H] and 
$1\sigma$ dispersion around the mean are indicated in each panel together with the 
number of stars in each region. The vertical dotted lines indicate the median [Fe/H].}
\label{fig:rgrad_mdf_spheroid2}
\end{figure*}

The analysis of the star counts around NGC~891 has shown the presence, in addition 
to the thick disk, of a spheroidal component with a de Vaucouleurs-like profile from 
$r \sim 0.5$~kpc to the edge of the survey at $r \sim 25$~kpc. This morphological 
component consists of an inner spheroid, prominent between $-4 \la X\la 6$~kpc and 
$-5 \la Z \la -2$~kpc, and the halo (see Fig. 3 of Paper III). The two-dimensional 
fit of the star number count distribution indicates that the spheroid becomes more flattened 
with distance, changing from $q=0.73 \pm 0.01$ in the inner parts to 
$q = 0.50 \pm 0.03$ in the outermost halo region probed. To sample the stellar 
populations of the inner spheroid, we select stars with $-6 < X< 6$~kpc, and 
$-5 < Z < -2$~kpc. To avoid contamination from the inner spheroid and/or the thick 
disk, halo stars are selected as those with $|Z|>6$~kpc and $X>6$~kpc. As in the 
case of the thick disk stellar population analysis, we restrict the sample to stellar 
density regions with less than 650 RGB stars per kpc$^2$.

The colour-magnitude diagrams of both inner spheroid and the halo are shown in the 
left panels of Fig.~\ref{fig:cmd_mdf_innersph} and Fig.~\ref{fig:cmd_mdf_halo}, respectively. 
The CMDs of both components look strikingly similar and, as for the thick disk, 
dominated by old stars, with RGB stars covering a wide range of colours, indicative 
of a mixture of low-to-intermediate metallicities. The MDFs of selected stars in 
each component are shown in the right panels of Fig.~\ref{fig:cmd_mdf_innersph} 
and Fig.~\ref{fig:cmd_mdf_halo}, respectively. The average metallicity of stars of
the inner spheroid is $-1.15$~dex, with a spread of $0.35$~dex and a median of $-1.08$~dex. 
For the halo, the average [Fe/H] is $-1.20$~dex, one sigma spread $0.34$~dex, and the median $-1.15$~dex.
The stellar populations dominating the two components of the spheroid surrounding 
the galaxy seem to be almost identical. This was remarked by \citet{ibata+09}, who 
found that inner spheroid and halo share similar colour and structure properties. 
Here we repeat their caveat: the inner most parts of the ``bulge", inside the $\sim 2$~kpc 
radius are almost impossible to probe with resolved stellar populations, while 
integrated light suffers from substantial extinction from the dust in the disk.

To investigate the spatial variations of stellar metallicity, we compute the 
median colour $(V-I)_{0,-3.5}$ and the median metallicity in elliptical rings, where 
we use different axis ratios for the inner spheroid and the halo. 
Fig.~\ref{fig:mdfgrad_halo} shows the the evolution of the median colour and 
metallicity as a function of minor axis distance. A modest metallicity gradient, 
i.e., $\sim 0.02$~dex/kpc, is present at about $3\sigma$ level. The scatter around 
the mean metallicity-distance relation is substantial however. 
The $1\sigma$ dispersion around the median metallicity is remarkably constant 
across the outer spheroid with $\sigma \sim 0.35$~dex. The presence of small-scale 
chemical substructures is evident from the large scatter around the mean relation.

Fig\,\ref{fig:rgrad_mdf_spheroid2} shows the spatial variation of the normalised,
extinction-corrected stellar metallicity distributions of the inner spheroid (left) 
and the halo (right). Stars have been selected in elliptical rings with the minor 
axis distances (b) indicated in each panel. The properties of the stellar content of 
the inner spheroid appear to be remarkably invariant over an extended range of 
distances along the minor axis. Beside the mild metallicity gradient, 
with average metallicity decreasing from about $-1.15$ to $-1.27$~dex from the inner
most regions to the outskirts (Fig.\,\ref{fig:mdfgrad_halo}), the properties of the stellar halo metallicity 
distribution appear to be the same over the surveyed area. No evidence that the 
contribution of the metal-poor peak of the stellar halo MDF increases at larger 
distances is present. No second stellar halo component is present in the halo of 
NGC~891, at least within the range of distance probed by our survey.

\section{Discussion}
\label{sec:discuss}

\begin{figure}
\resizebox{\hsize}{!}{
\includegraphics[angle=0]{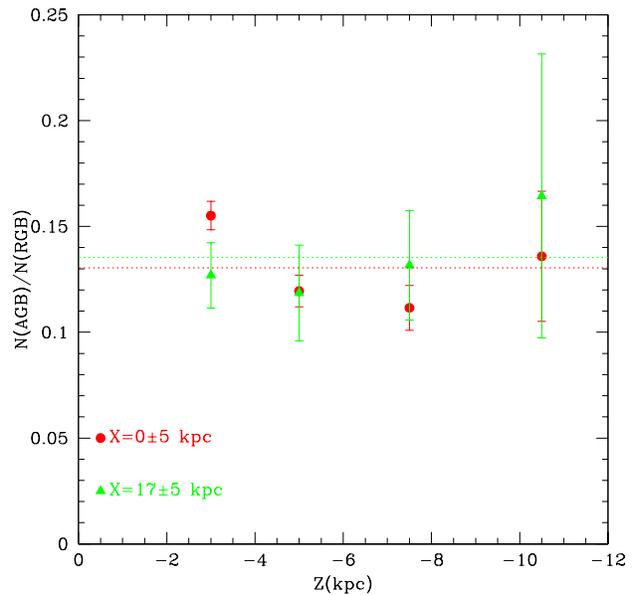}}
\caption{Ratio of AGB vs. RGB stars as a function of distance from the plane (Z), and 
for two different distances along the disk direction, at $X=17 \pm 5$~kpc (green 
triangles) and $X=0 \pm 5$~kpc (red dots).}
\label{fig:AGBrat}
\end{figure}

Stars brighter than the RGB tip can either be foreground contaminants, or old and 
metal-rich ([Fe/H]$\ga -0.6$~dex) TP-AGB stars, or intermediate-age TP-AGB, or blends 
of RGB tip stars. In the most inner fields, i.e., $Z<2$~kpc, most of bright stars are 
expected to be blends. While some blends are expected to be present in the fields 
with stellar density larger than $N(RGB)/\mathrm{kpc}^2 \ga 650$, the blends of two 
RGB tip stars that would mimic a bright AGB star is expected to be negligible in the 
regions with lower densities: at the distance of NGC~891, 1 kpc subtends about 420 
pixels, and there is only one bright red giant within 1 magnitude of the RGB tip 
for every 120 fainter RGB stars in an old stellar population with solar metallicity 
\citep{renzini98}. Judging by the lack of stars with metallicities above approximately
half solar in the probed field, the bulk of stars above the RGB tip are probable 
intermediate-age AGB stars. To investigate whether an age gradient is present across 
the thick disk and halo, we plot the ratio between AGB and RGB stars in 
Figure~\ref{fig:AGBrat}. Filled (red) dots show this ratio along the minor 
axis, while the (green) triangles are used to plot the ratio at X=17~kpc from the 
centre. In both cases the number ratio of AGB and RGB stars is computed for the 
same range of distances above the disk ($|\mathrm{Z}|$=10.5, 7.5, 5, and 3 kpc). AGB stars 
are selected between 0.2 and 1.0 magnitudes brighter than the RGB tip, while the RGB 
stars are selected between the RGB tip and 0.5 magnitudes fainter. The number ratio 
is fairly constant with radius and perpendicular to the disk plane. Accounting for 
the Poissonian errors ($1\sigma$ error-bars are plotted) the ratio remains constant 
at $\sim$ 0.13 across the surveyed thick disk and halo areas. This is strikingly similar 
to the measured ratios for three small edge-on galaxies and is consistent with an
old stellar population \citep{mould05}.
No significant bias is expected then when converting RGB photometry into metallicity.

Stars populating the halo and thick disk of NGC~891 are found to be predominantly 
old, similar to what is well established for the Milky Way halo \citep[e.g.][]{ryan+norris91}, 
and the thick disk \citep[e.g.][]{gilmore+reid83,fuhrmann04,bensby+03}. The vertical 
variations of the shape of the metallicity distributions of stars located at solar 
neighbourhood-like distances, where it is well determined in the case of the Milky 
Way, are significantly different in both galaxies however. This indicates that the 
mix of stellar populations that have been assembled to form the outskirts of the 
thick disks of both galaxies is different. Probably more massive or later accretions 
happened in NGC~891. A metal-poor thick disk and halo might be present but with a 
more extended component dominating the probed regions, or alternatively a large 
accretion (unidentified given the restricted extent of the area surveyed here) 
has polluted the stellar populations within the few kpc above the disk of NGC~891. 
A survey delivering a panoramic view of the outskirts of NGC~891 is needed. 
Whatever the exact driver(s) of the observed different vertical structures of the 
Milky Way and NGC~891, it is safe to conclude that the accretion histories of both 
galaxies have been substantially different.

The scale-heights of the Milky Way thin and thick disks are $\sim300$~pc, and 
$\sim900$~pc respectively \citep{juric+08}, and essentially all the stars within 
the $|\mathrm{Z}|>1.5$~kpc above the plane belong to the thick disk. NGC~891 has a larger 
thick disk, with a scale height of $1.44 \pm 0.03$~kpc and the scale length of 
$4.8\pm 0.1$~kpc \citep{ibata+09}. The thin and the thick disks of NGC~891 have similar 
radial scale lengths, with $h_R(thick)=4.8\pm0.1$~kpc, and $h_R(thin)=4.19\pm 0.01$~kpc.
The structure of the thick disk appears to be similar at different distances above the 
plane. What can this similarity tell us about the formation of the thick disk of NGC~891?

The metallicity distributions of the Galactic thin and thick disks overlap 
\citep[][and reference therein]{freeman+bh02}. It is still debated whether the thick disk 
metallicity distribution extends up to solar values, or at most to $\mathrm[Fe/H] \sim$0.2~dex 
\citep{bensby+07, fuhrmann08}. However, there is a discontinuity in the $\alpha$-element 
abundances between the thin and thick disk in the MW, which points to a disjoint 
formation histories for the two disks. The metallicity distribution of the thick disk 
of NGC~891 does not seem to extend to such high metallicity. However, our conclusion 
is limited due to possible bias in the inner regions due to blending. Observationally challenging 
spectroscopy of the low surface brightness regions of the NGC~891 thick disk is therefore needed 
to measure the $\alpha$-element abundances.


The very high stellar density and the presence of significant amounts of dust in 
the plane of the galaxy, prevent the analysis of the thin disk population in similar 
detail as is feasible for the thick disk and halo. The colour composite of our ACS images 
(Fig~\ref{N891_f1deep} in this paper and Fig~3 of Paper~III) shows the presence of 
blue, young stars and HII regions in the thin disk. Clearly the star formation is still 
on-going in the thin disk. The question is whether the thick disk may have formed from 
migrated thin disk stars, dispersed to higher orbits by some heating mechanism 
\citep{norris87}. The similar sizes of the thin and thick disk support the stellar 
diffusion formation scenario. This scenario is, however, not supported by the observed 
lack of vertical metallicity/colour gradient of thick disk stars. This is consistent 
with the finding of \citet{dalcanton+bernstein02, yoachim+dalcanton05} who have found 
that the scale lengths of thin and thick disks (derived from fitting of vertical surface 
brightness profiles) are not correlated.

Alternative scenarios for the thick disk formation include a slow pressure-supported 
collapse, violent chemical heating of the early thin disk by satellite accretion or 
violent relaxation of the galactic potential, direct accretion of thick disk material, 
and the rapid dissipational collapse \citep[e.g][]{gilmore+95}. In the slow dissipative 
disk formation scenario disk settling on time-scales longer than chemical enrichment 
timescale, would give rise to metallicity gradients of about 2~dex/kpc
\citep[e.g.][]{burkert+92}, which are much stronger than we observe in NGC~891. 
Furthermore, in that scenario the vertical gradient is expected to be present also 
in the inner regions, in contradiction with the observed properties. Our finding of
no vertical gradient in the inner thick disk regions, and no radial colour gradient, 
combined with the evidences of inhomogeneities seem to point to a combination of 
possible rapid dissipational collapse, or fast assembly of few gas-rich satellites 
at high redshift \citep[$z\sim 1.5-2$][]{brook+04b,brook+05}. Although the data point to 
a merger/accretion origin for the thick disk, it is difficult to disentangle models 
in which thick disk stars are accreted from those in which the stars form in situ 
further off the midplane during gas-rich mergers \citep{brook+04b}. Stars that formed 
in subhalos before being accreted are likely to have different properties than those 
that formed from accreted gas. Presumably, one could use detailed stellar kinematics, 
age and abundance information to distinguish between the two scenarios.

The stellar halo of NGC~891 shows a shallow metallicity gradient, i.e., $\sim0.02$~dex/kpc, 
with stars getting less chemically enriched in the outer regions. This is similar to what 
is found in the inner halo of M31~\citep{durrell+01}, and in the outskirts of the giant 
elliptical NGC~5128 \citep{rejkuba+05}. However, at distances larger than $\sim 12$~r$_{eff}$ 
from the centres of M31 and giant elliptical galaxy NGC~3379, the metal-poor halo was 
detected \citep{chapman+06, kalirai+06, harris+07b}. For NGC~891, the 12~r$_{eff}$ 
corresponds to about 21.5~kpc. The metal-poor halo, similar to those found in the MW and 
outer regions of M31, is then either missing, or we do not detect it at 13~r$_{eff}$ 
due to higher flattening of the halo and a more extended and/or less populated metal-poor 
component. This points again toward the need for a panoramic coverage of the outskirts 
of this analogue of the Milky Way to map the shape and the properties of its stellar 
halo.

\section{Summary \& Conclusions}

\label{conclusions}

We have derived an astrophotometric catalogue of 377320 stars detected in both $F606W$ 
and $F814W$ filters in three HST ACS fields in the north-eastern quadrant of NGC~891. 
A detailed description of the data reductions, completeness simulations, and photometric 
error analysis is presented. The final photometry is calibrated onto the ground based 
$VI$ Johnson-Cousins system.

The colour-magnitude diagrams are morphologically dominated by red giant branch 
stars, with no significant number of stars brighter than the classic old tip of 
the red giant branch. The number ratio of AGB stars to RGB stars is indicative 
of predominantly old stellar populations, and is roughly constant across the thick 
disk and the halo. 

The metallicity gradient of the thick disk population perpendicular to the plane 
of the galaxy is mild, amounting to $\Delta[Fe/H]/\Delta|Z| = 0.05 \pm 0.01$~kpc$^{-1}$, 
with bluer colours (lower metallicity) at higher distances from the plane. 
This is however fully dominated by the gradient in the outer regions of the thick 
disk, at distances larger than about 14~kpc from the centre along the major axis. 
The inner thick disk metallicity distribution is consistent with no gradient, 
similar to what is observed for Milky Way thick disk stars \citep{gilmore+95}, 
and suspected for other spiral galaxies from their vertical colour gradients.
In the radial direction data are consistent with no colour gradient, but with 
strong variations $1\sigma = 0.13$~mag around the mean colour of $(V-I)_{0,-3.5}=1.65$. 
The overall metallicity distribution of the thick disk stars peaks around 
$\mathrm{[Fe/H]}=-0.9$~dex, has the mean of $\mathrm{[Fe/H]}=-1.1$~dex with a
$1\sigma$ scatter of $0.34$~dex. 
The metallicity distribution functions of thick disk stars do not show any 
significant variation in either the vertical direction, or along the major axis.

A difference between the MW and NGC~891 is provided by 
the significantly different vertical variations of stellar metallicity distributions 
at solar circle-like locations. In NGC~891 there is a lack of a significant 
meal-poor component at high distances above the plane, i.e., $Z\ga 3$~kpc, that 
could be compared to the metal-poor MW halo. As concluded in Paper~I for regions
at $Z\sim 10$~kpc, the extra-planar stellar populations at lower heights from the 
disk of NGC~891 are more chemically enriched than those at similar locations in 
the MW. The presence of a significant scatter of thick disk stellar metallicities 
along the vertical and the radial directions might indicate that the observed 
difference could be due to a large accretion that has affected the properties of 
stars over the entire surveyed area. 

A metallicity gradient is detected for the spheroid changing from 
$\mathrm{[Fe/H]}=-1.15$~dex in the inner regions to $-1.27$~dex in
the outermost halo regions, with a large scatter around the mean metallicities of $\sim 0.35$~dex
throughout. 
The inner spheroid and the halo component up to about $12$~r$_{eff}$ shows
remarkable similarity in structure and stellar populations, with quite high average 
metallicity for the halo.

\section*{Acknowledgments} 
This work was based on observations with the NASA/ESA Hubble Space 
Telescope, obtained at the Space Telescope Science Institute, which 
is operated by the Association of Universities for Research in 
Astronomy, Inc.,under NASA contract NAS 5-26555. We thank Andy Dolphin for 
suggestions regarding DOLPHOT data reduction package, and Tom Oosterloo for
providing his high resolution deep HI map of the galaxy.

%
%
\bibliographystyle{mn2e}
\bibliography{/Users/mrejkuba/Work/publications/Article/mybiblio}

%
%

%
%

\bsp

\label{lastpage}

\end{document}

%% file: ngc891_paper2_tab01.tex
\begin{table*}
 \centering
 \begin{minipage}{140mm}
\caption{The observing log.}
\label{table:logs}
\begin{tabular}{@{}lrrlcc@{}}
\hline
Field Name & \multicolumn{1}{c}{RA}  & \multicolumn{1}{c}{DEC} & 
\multicolumn{1}{c}{Date} & \multicolumn{1}{c}{Filter} & \multicolumn{1}{c}{Exptime (sec)} \\
\hline
NGC891-HALO1 & 02:22:42.70 & +42:19:42.0 & 2003-02-19 & F606W &$3\times 824$\\
NGC891-HALO1-OFF1 & 02:22:42.74 & +42:19:42.0 & 2003-02-20 & F606W &$3\times 873$\\
NGC891-HALO1-OFF2 & 02:22:42.74 & +42:19:42.4 & 2003-02-20 & F606W &$3\times 873$\\
NGC891-HALO1 & 02:22:42.70 & +42:19:42.0 & 2003-02-18 & F814W &$3\times 824$\\
NGC891-HALO1-OFF1 & 02:22:42.74 & +42:19:42.0 & 2003-02-18 & F814W &$3\times 873$\\
NGC891-HALO1-OFF2 & 02:22:42.74 & +42:19:42.4 & 2003-02-18 & F814W &$3\times 873$\\
NGC891-HALO2 & 02:22:49.70 & +42:22:49.0 & 2003-02-17 & F606W &$3\times 824$\\
NGC891-HALO2-OFF1 & 02:22:49.74 & +42:22:49.0 & 2003-02-17 & F606W &$3\times 873$\\
NGC891-HALO2-OFF2 & 02:22:49.74 & +42:22:49.6 & 2003-02-17 & F606W &$3\times 873$\\
NGC891-HALO2 & 02:22:49.70 & +42:22:49.0 & 2003-02-16 & F814W &$3\times 824$\\
NGC891-HALO2-OFF1 & 02:22:49.74 & +42:22:49.0 & 2003-02-16 & F814W &$3\times 873$\\
NGC891-HALO2-OFF2 & 02:22:49.74 & +42:19:49.9 & 2003-02-16 & F814W &$3\times 873$\\
NGC891-HALO3 & 02:22:56.60 & +42:25:54.7 & 2003-02-17 & F606W &$3\times 824$\\
NGC891-HALO3-OFF1 & 02:22:56.64 & +42:22:54.7 & 2003-02-17 & F606W &$3\times 873$\\
NGC891-HALO3-OFF2 & 02:22:56.64 & +42:22:55.1 & 2003-02-17 & F606W &$3\times 873$\\
NGC891-HALO3 & 02:22:56.60 & +42:22:54.7 & 2003-02-17 & F814W &$3\times 824$\\
NGC891-HALO3-OFF1 & 02:22:56.64 & +42:22:54.7 & 2003-02-17 & F814W &$3\times 873$\\
NGC891-HALO3-OFF2 & 02:22:56.64 & +42:19:55.1 & 2003-02-17 & F814W &$3\times 873$\\
\hline
\end{tabular}
\end{minipage}
\end{table*}

%% file: ngc891_paper2_tab05.tex
\begin{table*}
 \centering
\caption{Statistics (number of stars N, mean, $1\sigma$ dispersion and median values) of 
the color and metallicity distributions of thick disk stars in
different bins along the minor (Z) and major (X) axis.}
\label{tab:vimdf_thick}
\begin{tabular}{@{}rrrrrrrrrrrr@{}}
\hline
\multicolumn{1}{c}{Z range}  & \multicolumn{1}{c}{X range} & 
\multicolumn{1}{c}{N$_{(V-I)}$} & \multicolumn{1}{c}{$(V-I)_{0,-3.5}$} & \multicolumn{1}{c}{$\sigma_{(V-I)}$} & \multicolumn{1}{c}{$(V-I)_{0,-3.5}$} &
\multicolumn{1}{c}{N$_{[Fe/H]}$} & \multicolumn{1}{c}{[Fe/H]} & \multicolumn{1}{c}{$\sigma_{[Fe/H]}$} & \multicolumn{1}{c}{[Fe/H]} \\
\multicolumn{1}{c}{kpc}  & \multicolumn{1}{c}{kpc} & 
\multicolumn{1}{c}{} & \multicolumn{1}{c}{mean mag} & \multicolumn{1}{c}{} & \multicolumn{1}{c}{median mag} &
\multicolumn{1}{c}{} & \multicolumn{1}{c}{mean dex} & \multicolumn{1}{c}{} & \multicolumn{1}{c}{median dex} \\
\hline
$-1>Z>-4$ & $ 7<X<9$ & 311	  & 1.59   & 0.30 & 1.57 &  552 & -1.17 & 0.35 & -1.11 \\
$-1>Z>-4$ & $ 9<X<11$ & 356	  & 1.61   & 0.31 & 1.59 &  625 & -1.19 & 0.38 & -1.11 \\
$-1>Z>-4$ & $14<X<16$ & 478	  & 1.67   & 0.33 & 1.66 &  944 & -1.05 & 0.30 & -0.99 \\
$-1>Z>-4$ & $16<X<18$ & 320	  & 1.66   & 0.32 & 1.62 &  654 & -1.04 & 0.31 & -0.96 \\
$-1>Z>-4$ & $18<X<20$ & 138	  & 1.71   & 0.30 & 1.67 &  297 & -1.05 & 0.32 & -0.99 \\
$-1>Z>-4$ & $20<X<25$ & 38	  & 1.58   & 0.27 & 1.58 &   67 & -1.19 & 0.39 & -1.15 \\
\hline
$-1.6>Z> -2.4$ & $ 7<X<  9$ &	0 &   $-$ &  $-$ &  $-$ &    0 &  $-$ &  $-$ &  $-$  \\ 
$-1.6>Z> -2.4$ & $ 9<X< 11$ &  65 & 1.55 & 0.32 & 1.53 &   82 & -1.28 & 0.44 & -1.15 \\ 
$-1.6>Z> -2.4$ & $14<X< 16$ & 175 & 1.68 & 0.31 & 1.68 &  343 & -1.02 & 0.28 & -0.97 \\ 
$-1.6>Z> -2.4$ & $16<X< 18$ &  96 & 1.70 & 0.34 & 1.69 &  212 & -1.01 & 0.27 & -0.93 \\
$-1.6>Z> -2.4$ & $18<X< 20$ &  55 & 1.73 & 0.29 & 1.72 &  105 & -1.05 & 0.30 & -0.99 \\ 
$-1.6>Z> -2.4$ & $20<X< 25$ &	6 & 1.51 & 0.20 & 1.61 &    6 & -0.86 & 0.23 & -0.81 \\ 
\hline
$-2.4>Z> -3.2$ & $ 7<X<  9$ & 120  & 1.61 & 0.30 & 1.59 & 214  & -1.15 & 0.33 & -1.12 \\ 
$-2.4>Z> -3.2$ & $ 9<X< 11$ & 187  & 1.62 & 0.32 & 1.60 & 335  & -1.20 & 0.39 & -1.11 \\
$-2.4>Z> -3.2$ & $14<X< 16$ &  75  & 1.71 & 0.31 & 1.71 & 145  & -1.02 & 0.28 & -0.95 \\
$-2.4>Z> -3.2$ & $16<X< 18$ &  53  & 1.65 & 0.28 & 1.60 &  84  & -1.06 & 0.34 & -0.96 \\
$-2.4>Z> -3.2$ & $18<X< 20$ &  24  & 1.71 & 0.25 & 1.69 &  57  & -1.13 & 0.36 & -1.06 \\ 
$-2.4>Z> -3.2$ & $20<X< 25$ &  14  & 1.58 & 0.38 & 1.58 &  26  & -1.23 & 0.42 & -1.15 \\ 
\hline
$-3.2>Z> -4.0$ & $ 7<X<  9$ & 191 & 1.58 & 0.30 & 1.55 & 338 & -1.18 & 0.36 & -1.11 \\ 
$-3.2>Z> -4.0$ & $ 9<X< 11$ & 104 & 1.62 & 0.29 & 1.60 & 208 & -1.15 & 0.33 & -1.09 \\
$-3.2>Z> -4.0$ & $14<X< 16$ &  47 & 1.69 & 0.40 & 1.65 & 100 & -1.13 & 0.35 & -1.05 \\
$-3.2>Z> -4.0$ & $16<X< 18$ &  24 & 1.53 & 0.34 & 1.49 &  67 & -1.13 & 0.37 & -1.04 \\
$-3.2>Z> -4.0$ & $18<X< 20$ &  22 & 1.52 & 0.21 & 1.55 &  43 & -1.06 & 0.32 & -0.97 \\ 
$-3.2>Z> -4.0$ & $20<X< 25$ &  18 & 1.60 & 0.20 & 1.61 &  35 & -1.22 & 0.37 & -1.20 \\ 
\hline
\end{tabular}
\end{table*}